\documentclass[journal]{IEEEtran}
\usepackage{multirow}
\usepackage{diagbox}
\usepackage{amssymb}
\usepackage{amsmath}
\usepackage{graphicx}
\usepackage{cite}
\usepackage{citesort}
\usepackage{subfigure}
\usepackage{graphicx,epstopdf}
\usepackage{epsfig}	
\usepackage{cite,graphicx,amsmath,amssymb}
\usepackage{comment}

\usepackage{amssymb}
\usepackage{amsmath}
\usepackage{cite}
\usepackage{url}
\usepackage{xcolor}
\usepackage{cite,graphicx,amsmath,amssymb}
\usepackage{subfigure}
\usepackage{citesort}
\usepackage{fancyhdr}
\usepackage{mdwmath}
\usepackage{mdwtab}
\usepackage{caption}
\usepackage{amsthm}
\usepackage{lipsum}

\newtheorem{theorem}{Theorem}

\newtheorem{lemma}{Lemma}

\newtheorem{corollary}{Corollary}

\newtheorem{remark}{Remark}  
\newtheorem{proposition}{Proposition}

\makeatletter
\def\ScaleIfNeeded{%
\ifdim\Gin@nat@width>\linewidth \linewidth \else \Gin@nat@width
\fi } \makeatother

\begin{document}

\title{\Huge{Modeling and Analysis of Two-Way Relay Non-Orthogonal Multiple Access Systems}}

\author{ Xinwei~Yue,~\IEEEmembership{Student Member,~IEEE,} Yuanwei\ Liu,~\IEEEmembership{Member,~IEEE,}
 Shaoli~Kang, Arumugam~Nallanathan,~\IEEEmembership{Fellow,~IEEE},
 and Yue Chen,~\IEEEmembership{Senior Member,~IEEE}

\thanks{X. Yue and S. Kang are with School of Electronic and Information Engineering, Beihang university, Beijing 100191,
China. S. Kang is also with State Key Laboratory of Wireless Mobile Communications, China Academy of Telecommunications Technology(CATT), Beijing 100094, China (email: xinwei$\_$yue@buaa.edu.cn, kangshaoli@catt.cn).}
\thanks{Y. Liu, A. Nallanathan and Y. Chen are with School of Electronic Engineering and Computer Science, Queen Mary University of London, London E1 4NS, UK (email: \{yuanwei.liu, a.nallanathan and yue.chen\}@qmul.ac.uk). Part of this work has been submitted to IEEE ICC 2018 \cite{Yue2018TWRNOMA}.}
}

\maketitle

\begin{abstract}
A two-way relay non-orthogonal multiple access (TWR-NOMA) system is investigated, where two groups of NOMA users exchange messages with the aid of one half-duplex (HD) decode-and-forward (DF) relay.
Since the signal-plus-interference-to-noise ratios (SINRs) of NOMA signals mainly depend on effective successive interference cancellation (SIC) schemes, imperfect SIC (ipSIC) and perfect SIC (pSIC) are taken into account. In order to characterize the performance of TWR-NOMA systems, we first derive closed-form expressions for both exact and asymptotic outage probabilities of NOMA users' signals with ipSIC/pSIC. Based on the derived results, the diversity order and throughput of the system are examined.
Then we study the ergodic rates of users' signals by providing the asymptotic analysis in high SNR regimes.
Lastly, numerical simulations are provided to verify the analytical results and show that: 1) TWR-NOMA is superior to TWR-OMA in terms of outage probability in low SNR regimes; 2) Due to the impact of interference signal (IS) at the relay, error floors and throughput ceilings exist in outage probabilities and ergodic rates for TWR-NOMA, respectively; and 3) In delay-limited transmission mode, TWR-NOMA with ipSIC and pSIC have almost the same energy efficiency. However, in delay-tolerant transmission mode, TWR-NOMA with pSIC is capable of achieving larger energy efficiency compared to TWR-NOMA with ipSIC.
\end{abstract}
\section{Introduction}
With the purpose to improve system throughput and spectrum efficiency, the fifth generation (5G) mobile communication networks are receiving a great deal of attention. The requirements of 5G networks mainly contain key performance indicator (KPI) improvement and support for new radio (NR) scenarios \cite{5Gscenarios}, including enhanced mobile broadband (eMBB), massive machine type communications (mMTC), and ultra-reliable and low latency communications (URLLC). Apart from crux technologies, such as massive multiple-input multiple-output (MIMO), millimeter wave and heterogeneous networks, the design of novel multiple access (MA) techniques is significant to make the contributions for 5G networks. Driven by these, non-orthogonal multiple access (NOMA) has been viewed as one of promising technologies to increase system capacity and user access \cite{Dai7263349}. The basic concept of NOMA is to superpose multiple users by sharing radio resources (i.e., time/frequencey/code) over different power levels \cite{Ding2017Mag,QinNOMA}. Then the desired signals are detected by exploiting the successive interference cancellation (SIC) \cite{Cover1991Elements}. More specifically, downlink multiuser superposition (MUST) transmission \cite{MUST1}, which is one of special case for NOMA has been researched for Long Term Evolution (LTE) in 3rd generation partnership project (3GPP) and approved as work item (WI) in radio access network (RAN) meeting.

Until now, point-to-point NOMA has been discussed extensively in many research contributions \cite{Ding6868214,Physical7812773,Haci7817823,Lv7870569}. In \cite{Ding6868214}, the authors have investigated the outage performance and ergodic rate of downlink NOMA with randomly deployed users by invoking stochastic geometry. Considering the secrecy issues of NOMA against external eavesdroppers, the authors in \cite{Physical7812773} investigated secrecy outage behaviors of NOMA in larger-scale networks for both single-antenna and multiple-antenna transmission scenarios. Explicit insights for understanding the asynchronous NOMA, a novel interference cancellation scheme was proposed in \cite{Haci7817823}, where the bit error rate and throughput performance were analyzed. By the virtue of available CSI, the performance of NOMA based multicast cognitive radio scheme (MCR-NOMA) was evaluated \cite{Lv7870569}, in which outage probability and diversity order are obtained for both secondary and primary networks.
Very recently, the application of cooperative communication \cite{laneman2004cooperative} to NOMA is an efficient way to offer enhanced spectrum efficiency and spatial diversity. Hence the integration of cooperative communication with NOMA has been widely discussed in many treaties \cite{Ding2014Cooperative,Kim2015Capacity,Liu7445146SWIPT,Men7454773}. Cooperative NOMA has been proposed in \cite{Ding2014Cooperative}, where the user with better channel condition acts as a decode-and-forward (DF) relay to forward information. Furthermore, in \cite{Kim2015Capacity}, the authors studied the ergodic rate of DF relay for a NOMA system.
With the objective of improving energy efficiency, the application of simultaneous wireless information and power transfer (SWIPT) to the nearby user was investigated where the locations of NOMA users were modeled by stochastic geometry \cite{Liu7445146SWIPT}.
Considering the impact of imperfect channel state information (CSI), the authors in \cite{Men7454773} investigated the performance of amplify-and-forward (AF) relay for downlink NOMA networks, where the exact and tight bounds of outage probability were derived. Moreover, in \cite{Men7752764}, the outage behavior and ergodic sum rate of NOMA for AF relay was analyzed under Nakagami-$m$ fading channels.
To further enhance spectrum efficiency, the performance of full-duplex (FD) cooperative NOMA was characterized in terms of outage probability \cite{Zhong7572025}.

Above existing treaties on cooperative NOMA are all based on one-way relay scheme, where the messages are delivered in only one direction, (i.e., from the BS to the relay or user destinations).
As a further advance, two-way relay (TWR) technique introduced in \cite{Shannon1961Two} has attracted remarkable interest as it is capable of boosting spectral efficiency. The basic idea of TWR systems is to exchange information between two nodes with the help of a relay, where AF or DF protocol can be employed. With the emphasis on user selection, in \cite{Jang5599948}, the authors analyzed the performance of multi-user TWR channels for half-duplex (HD) AF relays. By applying physical-layer network coding (PNC) schemes, the performance of two-way AF relay systems was investigated in terms of outage probability and sum rate \cite{Louie5403557}.
It was shown that two time slots PNC scheme achieves a higher sum rate compared to four time slot transmission mode.
In \cite{Hyadi7004056}, the authors studied the outage behaviors of DF relay with perfect and imperfect CSI conditions,
where a new relay selection scheme was proposed to reduce the complexity of TWR systems.
In terms of CSI and system state information, the system outage behavior was investigated for two-way full-duplex (FD) DF relay on different multi-user scheduling schemes \cite{Li7778253}.
In \cite{Song7064785}, the authors investigated the performance of multi-antenna TWR networks in which both AF and DF protocols are examined, respectively. Taking residual self-interference into account, the tradeoffs between the outage probability and ergodic rate were analyzed in \cite{Zhang7410116} for FD TWR systems. In addition, the authors in \cite{Sharma7463033} studied the performance of cooperative spectrum sharing by utilizing TWR over general fading channels. It was worth mentioning that the effective spectrum sharing is achieved by restraining additional cooperative diversity order.
\subsection{Motivations and Contributions}
While the aforementioned theoretical researches have laid a solid foundation for the understanding of NOMA and TWR techniques in wireless networks, the TWR-NOMA systems are far from being well understood. Obviously, the application of TWR to NOMA is a possible approach to improve the spectral efficiency of systems. To the best of our knowledge, there is no contributions to investigate the performance of TWR for NOMA systems.
Moreover, the above contributions for NOMA have been comprehensively studied under the assumption of perfect SIC (pSIC). In practical scenarios, there still exist several potential implementation issues with the use of SIC (i.e., complexity scaling and error propagation). More precisely, these unfavorable factors will lead to errors in decoding. Once an error occurs for carrying out SIC at the nearby user, the NOMA systems will suffer from the residual interference signal (IS). Hence it is significant to examine the detrimental impacts of imperfect SIC (ipSIC) for TWR-NOMA.
Motivated by these, we investigate the performance of TWR-NOMA with ipSIC/pSIC in terms of outage probability, ergodic rate and energy efficiency, where two groups of NOMA users exchange messages with the aid of a relay node using DF protocol.

The essential contributions of our paper are summarized as follows:
\begin{enumerate}
  \item
   We derive the closed-form expressions of outage probability for TWR-NOMA with ipSIC/pSIC. Based on the analytical results, we further derive the corresponding asymptotic outage probabilities and obtain the diversity orders. Additionally,
   we discuss the system throughput in delay-limited transmission mode.
   \item
   We show that the outage performance of TWR-NOMA is superior to TWR-OMA in the low signal-to-noise ratio (SNR) regime. We observe that due to the effect of IS at the relay, the outage probabilities for TWR-NOMA converge to error floors in the high SNR regime. We confirm that the use of pSIC is incapable of overcoming the zero diversity order for TWR-NOMA.
  \item
   We study the ergodic rate of users' signals for TWR-NOMA with ipSIC/pSIC. To gain more insights, we discuss one special case that when there is no IS between a pair of antennas at the relay. On the basis of results derived, we obtain the zero high SNR slopes for TWR-NOMA systems.
   We demonstrate that the ergodic rates for TWR-NOMA converge to throughput ceilings in high SNR regimes.
   \item
   We analyze the energy efficiency of TWR-NOMA with ipSIC/pSIC in both the delay-limited and tolerant transmission modes. We confirm that TWR-NOMA with ipSIC/pSIC in delay-limited transmission mode has almost the same energy efficiency.
   Furthermore, in delay-tolerant transmission mode, the energy efficiency of system with pSIC is higher than that of system with ipSIC.
\end{enumerate}

\subsection{Organization and Notation}
The remainder of this paper is organised as follows. In Section \ref{System Model}, the system mode for TWR-NOMA is introduced. In Section \ref{Section_III}, the analytical expressions for outage probability, diversity order and system throughput of TWR-NOMA are derived. Then the ergodic rates of users' signals for TWR-NOMA are investigated in Section \ref{Ergodic rate}. The system energy efficiency is evaluated in Section \ref{Energy Efficiency}. Analytical results and numerical simulations are presented in Section \ref{Numerical Results}, which is followed by our conclusions in Section \ref{Conclusion}.

The main notations of this paper is shown as follows:
$\mathbb{E}\{\cdot\}$ denotes expectation operation; ${f_X}\left(  \cdot  \right)$ and ${F_X}\left(  \cdot  \right)$ denote the probability density function (PDF) and the cumulative distribution function (CDF) of a random variable $X$.

\section{System Model}\label{System Model}
\subsection{System Description}
\begin{figure}[t!]
    \begin{center}
        \includegraphics[width=3.5in,  height=1.7in]{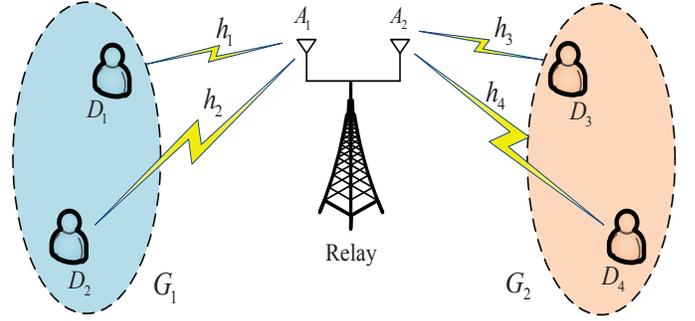}
        \caption{An illustration of TWR-NOMA systems, in which two groups of users exchange messages with the aid of one relay node.}
        \label{System model}
    \end{center}
\end{figure}

We focus our attentions on a two-way relay NOMA communication scenario which consists of one relay $R$, two pairs of NOMA users ${G_1} = \left\{ {{D_1},{D_2}} \right\}$ and ${G_2} =\left\{ {{D_3},{D_4}} \right\}$\footnote{The geographical dimensions of clusters $G_1$ and $G_2$ are to ensure that there is a certain distance difference from distant user and nearby user to $R$.}. To reduce the complexity of systems, many research contributions on NOMA have been proposed to pair two users for the application of NOMA protocol\footnote{Note that increasing the number of paired users, i,e,. $N$ pairs of users, will not affect the performance of TWR-NOMA system. It is worth pointing that within each group, superposition coding and SIC are employed, and across the groups, transmissions are orthogonal.} \cite{Pairing7273963,Ding7482785}.
As shown in Fig.~\ref{System model}, we assume that $D_{1}$ and $D_{3}$ are the nearby users in groups ${G_1}$ and ${G_2}$, respectively, while $D_{2}$ and $D_{4}$ are the distant users in groups ${G_1}$ and ${G_2}$, respectively. It is worth noting that the nearby user and distant user are distinguished based on the distance from the users to $R$ \cite{Yue8026173}. For example, $D_1$ and $D_3$ are near to $R$, while $D_2$ and $D_4$ are far away from $R$.
The exchange of information between user groups $G_{1}$ and $G_{2}$ is facilitated via the assistance of a decode-and-forward (DF) relay with two antennas, namely $A_1$ and $A_2$\footnote{For the practical scenario, we can assume that the relay is located on a mountain, where the user nodes on both sides of the mountain are capable of exchanging the information between each other.}.
User nodes are equipped with single antenna. 
In practical communication process, the complexity of DF protocol is too high to implement. To facilitate analysis, we focus our attention on a idealized DF protocol, where $R$ is capable of decoding the users' information correctly. Relaxing this idealized assumption can make system mode close to the practical scenario, but this is beyond the scope of this treatise. Additionally, to evaluate the impact of error propagation on TWR-NOMA, ipSIC operation is employed at relay $R$ and nearby users.
It is assumed that the direct links between two pairs of users are inexistent due to the effect of strong shadowing. Without loss of generality, all the wireless channels are modeled to be independent quasi-static block Rayleigh fading channels and disturbed by additive white Gaussian noise with mean power $N_{0}$. Furthermore, ${h_{1}}$, ${h_{2}}$, ${h_{3}}$ and ${h_{4}}$ are denoted as the complex channel coefficient of $D_{1}  \leftrightarrow R$, $D_{2} \leftrightarrow R$, $D_3 \leftrightarrow R$ and $D_4 \leftrightarrow R$ links, respectively. We assume that the channels from user nodes to $R$ and the channels from $R$ to user nodes are reciprocal. In other words, the channels from user nodes to $R$ have the same fading impact as the channels from $R$ to the user nodes \cite{Cui6812188,Zhang7410116,Chen}.
The channel power gains ${|h_{1}}|^2$, ${|h_{2}}|^2$, ${|h_{3}}|^2$ and ${|h_{4}}|^2$ are assumed to be exponentially distributed random variables (RVs) with the parameters $ \Omega_{i}$, $\emph{i}\in\{1,2,3,4\}$, respectively.
Note that the perfect CSIs of NOMA users are available at $R$ for signal detection.


\subsection{Signal Model}
During the first slot, the pair of NOMA users in $G_1$ transmit the signals to $R$ just as uplink NOMA.
Since $R$ is equipped with two antennas, when $R$ receives the signals from the pair of users in $G_1$, it will suffer from
interference signals from the pair of users in $G_2$. More precisely, the observation at $R$ for $A_{1}$ is given by
\begin{align}\label{the expression of R for A1}
{y_{{R_{{A_1}}}}} = {h_1}\sqrt {{a_1}P_u} {x_1} + {h_2}\sqrt {{a_2}P_u} {x_2} + {{\varpi _1}} {I_{{R_{{A_2}}}}} + {n_{{R_{{A_1}}}}},
\end{align}
where ${I_{{R_{{A_2}}}}} $ denotes IS from $A_2$ with ${I_{{R_{{A_2}}}}} = ({h_3}\sqrt {{a_3}P_u} {x_3} + {h_4}\sqrt{{a_4}P_u} {x_4})$. ${\varpi _1} \in \left[ {0,1} \right]$ denotes the impact levels of IS at $R$.
$P_u$ is the transmission power at user nodes.
$x_1$, $x_{2}$ and $x_{3}$, $x_{4}$ are the signals of $D_{1}$, $D_{2}$ and $D_{3}$, $D_{4}$, respectively, i.e,
$\mathbb{E}\{x_{1}^2\}= \mathbb{E}\{x_{2}^2\}=\mathbb{E}\{x_{3}^2\}=\mathbb{E}\{x_{4}^2\}=1$. $a_1$, $a_{2}$ and $a_{3}$, $a_{4}$ are
the corresponding power allocation coefficients. Note that the efficient uplink power control is capable of enhancing the performance of the systems considered, which is beyond the scope of this paper. ${n_{R_{A_{j}}}}$ denotes the Gaussian noise at $R$ for $A_j$, $j \in \left\{ {1,2}
\right\}$.

Similarly, when $R$ receives the signals from the pair of users in $G_2$, it will suffer from interference signals from the pair of users in $G_1$ as well and then the observation at $R$ is given by
\begin{align}\label{the expression of R for A2}
{y_{{R_{{A_2}}}}} = {h_3}\sqrt {{a_3}P_u} {x_3} + {h_4}\sqrt {{a_4}P_u} {x_4} + {{\varpi _1}} {I_{{R_{{A_1}}}}} + {n_{{R_{{A_2}}}}},
\end{align}
where ${I_{{R_{{A_1}}}}} $ denotes the interference signals from $A_1$ with ${I_{{R_{{A_1}}}}} = ({h_1}\sqrt {{a_1}P_u} {x_1} + {h_2}\sqrt
 {{a_2}P_u} {x_2})$.

Applying the NOMA protocol, $R$ first decodes $D_l$'s information $x_l$ by the virtue of treating $x_t$ as IS. Hence
the received signal-to-interference-plus-noise ratio (SINR) at $R$ to detect $x_l$ is given by
\begin{align}\label{the expression SINR of R to detect x1 or x3}
{\gamma _{R \to {x_l}}} = \frac{{\rho {{\left| {{h_l}} \right|}^2}{a_l}}}{{\rho {{\left| {{h_t}} \right|}^2}{a_t} + \rho {\varpi _1}({{\left| {{h_k}} \right|}^2}{a_k} + {{\left| {{h_r}} \right|}^2}{a_r}) + 1}},
\end{align}
where $\rho  = \frac{{{P_u}}}{{{N_0}}}$ denotes the transmit signal-to-noise ratio (SNR), $\left( {l,k} \right) \in \left\{ {\left( {1,3} \right),\left( {3,1} \right)} \right\}$, $\left( {t,r} \right) \in \left\{ {\left( {2,4} \right),\left( {4,2} \right)} \right\}$.

After SIC is carried out at $R$ for detecting $x_l$, the received SINR at $R$ to detect $x_t$ is given by
\begin{align}\label{the expression SINR of R to detect x2 or x4}
{\gamma _{R \to {x_t}}} = \frac{{\rho {{\left| {{h_t}} \right|}^2}{a_t}}}{{\varepsilon \rho {{\left| g \right|}^2} + \rho {\varpi _1}({{\left| {{h_k}} \right|}^2}{a_k} + {{\left| {{h_r}} \right|}^2}{a_r}) + 1}},
\end{align}
where $\varepsilon = 0$ and $\varepsilon = 1$ denote the pSIC and ipSIC employed at $R$, respectively. Due to the impact of ipSIC, the residual IS is modeled as Rayleigh fading channels \cite{RelaySharing7819537} denoted as $g$ with zero mean and variance ${{\Omega _I}}$.

In the second slot, the information is exchanged between $G_1$ and $G_2$ by the virtue of $R$. Therefore, just like the downlink NOMA, $R$
transmits the superposed signals $( {\sqrt {{b_1}P_r} {x_1} + \sqrt {{b_2}P_r} {x_2}} )$ and $( {\sqrt {{b_3}P_r} {x_3} + \sqrt {{b_4}P_r} {x_4}} )$ to $G_2$ and $G_1$ by $A_2$ and $A_1$, respectively.
$b_1$ and $b_{2}$ denote the power allocation coefficients of $D_1$ and $D_2$, while $b_{3}$ and $b_{4}$ are the corresponding power allocation coefficients of $D_3$ and $D_4$, respectively.
\textcolor[rgb]{0.00,0.00,1.00}{$P_{r}$ is the transmission power at $R$ and we assume $P_{u}= P_{r}$.} In particular, to ensure the fairness between users in $G_1$ and $G_2$, a higher power should be allocated to the distant user who has the worse channel condition. Hence we assume that ${b_2} > {b_1}$ with $b_{1} +b_{2} = 1$ and ${b_4} > {b_3}$ with $b_{3} +b_{4} = 1$. Note that the fixed power allocation coefficients for two groups' NOMA users are considered. Relaxing this assumption will further improve the performance of systems and should be concluded in our future work.

According to NOMA protocol, SIC is employed and the received SINR at $D_k$ to detect $x_t$ is given by
\begin{align}\label{the SINR expression for D1 or D3 to detect x4 or x2}
{\gamma _{{D_k} \to {x_t}}} = \frac{{\rho {{\left| {{h_k}} \right|}^2}{b_t}}}{{\rho {{\left| {{h_k}} \right|}^2}{b_l} + \rho {\varpi _2}{{\left| {{h_k}} \right|}^2} + 1}},
\end{align}
where ${\varpi _2} \in \left[ {0,1} \right]$ denotes the impact level of IS at the user nodes. Then $D_k$ detects $x_l$ and gives the corresponding SINR as follows:
\begin{align}\label{the SINR expression for D1 or D3 to detect its own information}
{\gamma _{{D_k} \to {x_l}}} = \frac{{\rho {{\left| {{h_k}} \right|}^2}{b_l}}}{{\varepsilon \rho {{\left| g \right|}^2} + \rho {\varpi _2}{{\left| {{h_k}} \right|}^2} + 1}}.
\end{align}
Furthermore, the received SINR at $D_r$ to detect $x_t$ can be given by
\begin{align}\label{the SINR expression for D2 or D4}
{\gamma _{{D_r} \to {x_t}}} = \frac{{\rho {{\left| {{h_r}} \right|}^2}{b_t}}}{{\rho {{\left| {{h_r}} \right|}^2}{b_l} + \rho {\varpi _2}{{\left| {{h_r}} \right|}^2} + 1}}.
\end{align}

From above process, the exchange of information is achieved between the NOMA users for $G_1$ and $G_2$.
More specifically, the signal ${x_1}$ of ${D_1}$ is exchanged with the signal ${x_3}$ of ${D_3}$. Furthermore, the signal ${x_2}$ of ${D_2}$ is exchanged with the signal ${x_4}$ of ${D_4}$.

\section{Outage Probability}\label{Section_III}
In this section, the performance of TWR-NOMA is characterized in terms of outage probability. Due to the channel's reciprocity, the outage probability of $x_l$ and $x_t$ are provided in detail in the following part.
\subsubsection{Outage Probability of $x_{l}$}\label{the expression of outage for x1}
In TWR-NOMA system, the outage events of $x_l$ are explained as: i) $R$ cannot decode $x_l$ correctly; ii) The information $x_t$ cannot be detected by $D_k$; and iii) $D_{k}$ cannot detect $x_l$, while $D_{k}$ can first decode $x_t$ successfully. To simplify the analysis, the complementary events of $x_l$ are employed to express its outage probability.
As a consequence, the outage probability of $x_{l}$ with ipSIC for TWR-NOMA system can be given by
\begin{align}\label{OP expression for x1}
 P_{{x_l}}^{ipSIC} =& 1 - \Pr \left( {{\gamma _{R \to {x_l}}} > {\gamma _{t{h_l}}}} \right) \nonumber \\
  &\times \Pr \left( {{\gamma _{{D_k} \to {x_t}}} > {\gamma _{t{h_t}}},{\gamma _{{D_k} \to {x_l}}} > {\gamma _{t{h_l}}}} \right)  ,
\end{align}
where $\varepsilon = 1$, ${\varpi _1} \in \left[ {0,1} \right]$ and ${\varpi _2} \in \left[ {0,1} \right]$. ${\gamma _{t{h_l}}} = {2^{2{R_l}}} - 1$ with $R_{l}$ being the target rate
at $D_{k}$ to detect $x_{l}$ and ${\gamma _{t{h_t}}} = {2^{2{R_t}}} - 1$ with $R_{t}$ being the target rate at $D_{k}$ to detect $x_{t}$.

The following theorem provides the outage probability of $x_{l}$ for TWR-NOMA.
\begin{theorem} \label{theorem:1 the outage of x1}
The closed-form expression for the outage probability of $x_{l}$ for TWR-NOMA with ipSIC is given by
\begin{align}\label{OP derived for x1}
&P_{{x_l}}^{ipSIC} = 1 - {e^{ - \frac{\beta_l }{{{\Omega _l}}}}}\prod\limits_{i = 1}^3 {{\lambda _i}} \left( {\frac{{{\Phi _1}{\Omega _l}}}{{{\Omega _l}{\lambda _1}{\rm{ + }}\beta_l }} - \frac{{{\Phi _2}{\Omega _l}}}{{{\Omega _l}{\lambda _2}{\rm{ + }}\beta_l }}} \right. \nonumber \\
 &\left. { + \frac{{{\Phi _3}{\Omega _l}}}{{{\Omega _l}{\lambda _3}{\rm{ + }}\beta_l }}} \right)\left( {{e^{ - \frac{\theta_l }{{{\Omega _k}}}}} - \frac{{\varepsilon \tau_l  \rho {\Omega _I}}}{{{\Omega _k} + \varepsilon \rho \tau_l {\Omega _I}}}{e^{ - \frac{{\theta_l \left( {{\Omega _k} + \varepsilon\rho \tau_l  {\Omega _I}} \right)}}{{\varepsilon \tau_l  \rho {\Omega _I}{\Omega _k}}}{\rm{ + }}\frac{1}{{\varepsilon \rho {\Omega _I}}}}}} \right)  ,
\end{align}
where $\varepsilon  = 1$. ${\lambda _1}{\rm{ = }}\frac{1}{{\rho {a_t}{\Omega _t}}}$, ${\lambda _2}{\rm{ = }}\frac{1}{{\rho {\varpi _1}{a_k}{\Omega _k}}}$ and ${\lambda _3}{\rm{ = }}\frac{1}{{\rho {\varpi _1}{a_r}{\Omega _r}}}$. $\beta_l {\rm{ = }}\frac{{{\gamma _{t{h_l}}}}}{{\rho {a_l}}}$. ${\Phi _1}{\rm{ = }}\frac{1}{{\left( {{\lambda _2} - {\lambda _1}} \right)\left( {{\lambda _3} - {\lambda _1}} \right)}}$,${\Phi _2}{\rm{ = }}\frac{1}{{\left( {{\lambda _3} - {\lambda _2}} \right)\left( {{\lambda _2} - {\lambda _1}} \right)}}$ and ${\Phi _3}{\rm{ = }}\frac{1}{{\left( {{\lambda _3} - {\lambda _1}} \right)\left( {{\lambda _3} - {\lambda _2}} \right)}}$. $\theta_l  \buildrel \Delta \over
 = \max \left( {\tau_l ,\xi_t } \right)$.
$\tau_l {\rm{ = }}\frac{{{\gamma _{t{h_l}}}}}{{\rho \left( {{b_l} - {\varpi _2}{\gamma _{t{h_l}}}} \right)}}$ with ${b_l} > {\varpi _2}{\gamma _{t{h_l}}}$ and $\xi_t {\rm{ = }}\frac{{{\gamma _{t{h_t}}}}}{{\rho \left( {{b_t} - {b_l}{\gamma _{t{h_t}}} - {\varpi _2}{\gamma _{t{h_t}}}} \right)}}$ with ${b_t} > \left( {{b_l} + {\varpi _2}} \right){\gamma _{t{h_t}}}$.
\begin{proof}
See Appendix~A.
\end{proof}

\end{theorem}
\begin{corollary}
Based on \eqref{OP derived for x1}, for the special case $\varepsilon = 0$,  the outage probability of $x_{1}$ for TWR-NOMA with pSIC is given by
\begin{align}\label{corollary1 derived for x1 with perfect SIC}
 P_{{x_l}}^{pSIC} =& 1 - {e^{ - \frac{\beta_l }{{{\Omega _l}}} - \frac{\theta_l }{{{\Omega _k}}}}}\prod\limits_{i = 1}^3 {{\lambda _i}} \left( {\frac{{{\Phi _1}{\Omega _l}}}{{{\Omega _l}{\lambda _1}{\rm{ + }}\beta_l }} - \frac{{{\Phi _2}{\Omega _l}}}{{{\Omega _l}{\lambda _2}{\rm{ + }}\beta_l }}} \right. \nonumber \\
 & \left. { + \frac{{{\Phi _3}{\Omega _l}}}{{{\Omega _l}{\lambda _3}{\rm{ + }}\beta_l }}} \right)  .
\end{align}
\end{corollary}
\subsubsection{Outage Probability of $x_{t}$}\label{the expression of outage for x2}
Based on NOMA principle, the complementary events of outage for $x_t$ have the following cases. One of the cases is that $R$ can first
decode the information $x_l$ and then detect $x_t$. Another case is that either of $D_{k}$ and $D_{r}$ can detect $x_t$ successfully.
Hence the outage probability of $x_{t}$ can be expressed as
\begin{align}\label{OP expression for x2}
 P_{{x_t}}^{ipSIC} =& 1 - \Pr \left( {{\gamma _{R \to {x_t}}} > {\gamma _{t{h_t}}},{\gamma _{R \to {x_l}}} > {\gamma _{t{h_l}}}} \right) \nonumber \\
 & \times \Pr \left( {{\gamma _{{D_k} \to {x_t}}} > {\gamma _{t{h_t}}}} \right)\Pr\left( {{\gamma _{{D_r} \to {x_t}}} > {\gamma _{t{h_t}}}} \right) ,
\end{align}
where $\varepsilon = 1$, ${\varpi _1} \in \left[ {0,1} \right]$ and ${\varpi _2} \in \left[ {0,1} \right]$.

The following theorem provides the outage probability of $x_{t}$ for TWR-NOMA.
\begin{theorem} \label{theorem:2 the outage of x2}
The closed-form expression for the outage probability of $x_{t}$ with ipSIC is given by
\begin{align}\label{OP derived for x2}
 &P_{{x_t}}^{ipSIC} = 1 - \frac{{{e^{ - \frac{\beta_l }{{{\Omega _l}}} - {\beta_t}\varphi_t  - \frac{\xi }{{{\Omega _k}}} - \frac{\xi }{{{\Omega _r}}}}}}}{{\varphi_t {\Omega _t}\left( {1 + \varepsilon {\beta_t} \rho \varphi_t {\Omega _I}} \right)\left( {\lambda _2^{'} - \lambda _1^{'}} \right)}}\prod\limits_{i = 1}^2 {\lambda _i^{'}}  \nonumber\\
 &  \times \left( {\frac{{{\Omega _l}}}{{\beta_l  + {\beta_t}{\Omega _l}\varphi_t  + {\Omega _l}\lambda _1^{'}}} - \frac{{{\Omega _l}}}{{\beta_l  + {\beta_t}{\Omega _l}\varphi_t  + {\Omega _l}\lambda _2^{'}}}} \right),
\end{align}
where $\varepsilon = 1$. $\lambda _1^{'}{\rm{ = }}\frac{1}{{\rho {\varpi _1}{a_k}{\Omega _k}}}$ and $\lambda _2^{'}{\rm{ = }}\frac{1}{{\rho {\varpi _1}{a_r}{\Omega _r}}}$. ${\beta_t} = \frac{{{\gamma _{t{h_t}}}}}{{\rho {a_t}}}$, $\varphi_t  = \frac{{{\Omega _l}+ \rho \beta_l {a_t}{\Omega _t}}}{{{\Omega _l}{\Omega _t}}}$.
\begin{proof}
See Appendix~B.
\end{proof}
\end{theorem}

\begin{corollary}
For the special case, substituting $\varepsilon  = 0$ into \eqref{OP derived for x2}, the outage probability of $x_{2}$ for TWR-NOMA with pSIC is
given by
\begin{align}\label{corollary2 derived for x2 with perfect SIC}
 &P_{{x_t}}^{pSIC} = 1 - \frac{{{e^{ - \frac{\beta_l }{{{\Omega _l}}} - {\beta_t}\varphi_t  - \frac{\xi }{{{\Omega _k}}} - \frac{\xi }{{{\Omega _r}}}}}}}{{\varphi_t {\Omega _t}\left( {\lambda _2^{'} - \lambda _1^{'}} \right)}}\prod\limits_{i = 1}^2 {\lambda _i^{'}} \nonumber \\
  & \times \left( {\frac{{{\Omega _l}}}{{\beta_l  + {\beta_t}{\Omega _l}\varphi_t  + {\Omega _l}\lambda _1^{'}}} - \frac{{{\Omega _l}}}{{\beta_l  + {\beta_t}{\Omega _l}\varphi_t  + {\Omega _l}\lambda _2^{'}}}} \right) .
\end{align}
\end{corollary}
\subsubsection{Diversity Order Analysis}
In order to gain deeper insights for TWR-NOMA systems, the asymptotic analysis are presented in high SNR regimes based on the derived outage probabilities. The diversity order is defined as \cite{Liu2016TVT}
\begin{align}\label{diversity order}
d =  - \mathop {\lim }\limits_{\rho  \to \infty } \frac{{\log \left( {P_{x_i}^\infty \left( \rho  \right)} \right)}}{{\log \rho }},
\end{align}
where ${P_{x_i}^\infty }$ denotes the asymptotic outage probability of $x_{i}$.

\begin{proposition}\label{proposition:diversity total for x_l}
Based on the analytical results in \eqref{OP derived for x1} and \eqref{corollary1 derived for x1 with perfect SIC}, when $\rho  \to \infty $, the asymptotic outage probabilities of $x_l$ for ipSIC/pSIC with ${e^{ - x}}\approx  1 - x$ are given by
\begin{align}\label{the asymptotic OP of x1 with ipSIC}
 &P_{{x_l},\infty }^{ipSIC} = 1 - \prod\limits_{i = 1}^3 {{\lambda _i}} \left( {\frac{{{\Phi _1}{\Omega _l}}}{{{\Omega _l}{\lambda _1}{\rm{ + }}{\beta _l}}} - \frac{{{\Phi _2}{\Omega _l}}}{{{\Omega _l}{\lambda _2}{\rm{ + }}{\beta _l}}} + \frac{{{\Phi _3}{\Omega _l}}}{{{\Omega _l}{\lambda _3}{\rm{ + }}{\beta _l}}}} \right) \nonumber \\
  &\times \left[ {1 - \frac{{{\theta _l}}}{{{\Omega _k}}} - \frac{{\varepsilon \tau_{l} \rho {\Omega _I}}}{{{\Omega _k} + \varepsilon \rho \tau_{l} {\Omega _I}}}\left( {1 - \frac{{{\theta _l}\left( {{\Omega _k} + \varepsilon \tau_{l} \rho {\Omega _I}} \right)}}{{\varepsilon \rho \tau_{l}  {\Omega _I}{\Omega _k}}}} \right)} \right] ,
\end{align}
and
\begin{align}\label{the asymptotic OP of x1 with pSIC}
P_{{x_l},\infty }^{pSIC} = 1 - \prod\limits_{i = 1}^3 {{\lambda _i}} \left( {\frac{{{\Phi _1}{\Omega _l}}}{{{\Omega _l}{\lambda _1}{\rm{ + }}{\beta _l}}} - \frac{{{\Phi _2}{\Omega _l}}}{{{\Omega _l}{\lambda _2}{\rm{ + }}{\beta _l}}} + \frac{{{\Phi _3}{\Omega _l}}}{{{\Omega _l}{\lambda _3}{\rm{ + }}{\beta _l}}}} \right),
\end{align}
respectively. Substituting \eqref{the asymptotic OP of x1 with ipSIC} and \eqref{the asymptotic OP of x1 with pSIC} into \eqref{diversity order}, the diversity orders of $x_l$ with ipSIC/pSIC are equal to zeros.
\end{proposition}

\begin{remark}\label{remark:1}
An important conclusion from above analysis is that due to impact of residual interference, the diversity order of $x_{l}$ with the use of ipSIC is zero. Additionally, the communication process of the first slot similar to uplink NOMA, even though under the condition of pSIC, diversity order is equal to zero as well for $x_{l}$. As can be observed that there are error floors for $x_{l}$ with ipSIC/pSIC.
\end{remark}
\begin{proposition}\label{proposition:diversity total for x_l}
Similar to the resolving process of $x_l$, the asymptotic outage probabilities of $x_t$ with ipSIC/pSIC in high SNR regimes are given by
\begin{align}\label{the asymptotic OP of x2 with ipSIC}
 &P_{{x_t},\infty }^{ipSIC} = 1 - \frac{{\lambda _1^{'}\lambda _2^{'}}}{{{\varphi _t}{\Omega _t}\left( {1 + \varepsilon \rho {\beta _t}{\varphi _t}{\Omega _I}} \right)\left( {\lambda _2^{'} - \lambda _1^{'}} \right)}} \nonumber \\
  &\times \left( {\frac{{{\Omega _l}}}{{{\beta _l} + {\beta _t}{\Omega _1}{\varphi _t} + {\Omega _l}\lambda _1^{'}}} - \frac{{{\Omega _l}}}{{{\beta _l} + {\beta _t}{\Omega _1}{\varphi _t} + {\Omega _l}\lambda _2^{'}}}} \right)  ,
 \end{align}
and
\begin{align}\label{the asymptotic OP of x2 with pSIC}
 &P_{{x_t},\infty }^{pSIC} = 1 - \frac{{\lambda _1^{'}\lambda _2^{'}}}{{{\varphi _t}{\Omega _t}\left( {\lambda _2^{'} - \lambda _1^{'}} \right)}}\nonumber \\
  & \times \left( {\frac{{{\Omega _l}}}{{{\beta _l} + {\beta _t}{\Omega _1}{\varphi _t} + {\Omega _l}\lambda _1^{'}}} - \frac{{{\Omega _l}}}{{{\beta _l} + {\beta _t}{\Omega _l}{\varphi _t} + {\Omega _l}\lambda _2^{'}}}} \right) ,
\end{align}
respectively. Substituting \eqref{the asymptotic OP of x2 with ipSIC} and \eqref{the asymptotic OP of x2 with pSIC} into \eqref{diversity order}, the diversity orders of $x_{t}$ for both ipSIC and pSIC are zeros.
\end{proposition}
\begin{remark}\label{remark:2}
Based on above analytical results of $x_{l}$, the diversity orders of $x_{t}$ with ipSIC/pSIC are also equal to zeros.
This is because residual interference is existent in the total communication process.
\end{remark}
\subsubsection{Throughput Analysis}\label{Seciton-A-3}
In delay-limited transmission scenario, the BS transmits message to users at a fixed rate, where system throughput will be subject to wireless fading channels. Hence the corresponding throughput of TWR-NOMA with ipSIC/pSIC is calculated as \cite{Liu7445146SWIPT,Nasir6552840}
\begin{align}\label{delay-limited throughput}
 R_{dl}^{\psi} =& \left( {1 - P_{{x_1}}^{\psi} } \right){R_{{x_1}}} + \left( {1 - P_{{x_2}}^{\psi} } \right){R_{{x_2}}} \nonumber \\
  &+ \left( {1 - P_{{x_3}}^{\psi} } \right){R_{{x_3}}} + \left( {1 - P_{{x_4}}^{\psi} } \right){R_{{x_4}}},
\end{align}
where $\psi  \in \left( {ipSIC,pSIC} \right)$. $P_{{x_1}}^{\psi}$ and $P_{{x_3}}^{\psi}$ with ipSIC/pSIC can be obtained from \eqref{OP derived for x1} and  \eqref{corollary1 derived for x1 with perfect SIC}, respectively, while
$P_{{x_2}}^{\psi}$ and $P_{{x_4}}^{\psi}$ with ipSIC/pSIC can be obtained from  \eqref{OP derived for x2} and \eqref{corollary2 derived for x2 with perfect SIC}, respectively.
\section{Ergodic rate}\label{Ergodic rate}
In this section, the ergodic rate of TWR-NOMA is investigated for considering the influence of signal's channel fading to target rate.
\setcounter{subsubsection}{0}
\subsubsection{Ergodic Rate of $x_{l}$}\label{Ergodic Rate of x1}
Since $x_{l}$ can be detected at the relay as well as at $D_k$ successfully. By the virtue of \eqref{the expression SINR of R to detect x1 or x3} and \eqref{the SINR expression for D1 or D3 to detect its own information}, the achievable rate of $x_l$ for TWR-NOMA is written as ${R_{{x_l}}} = \frac{1}{2}\log \left( {1 + \min \left( {{\gamma _{R \to {x_l}}},{\gamma _{{D_k} \to {x_l}}}} \right)} \right)$. In order to further calculate the ergodic rate of $x_l$, using $X = \min \left( {{\gamma _{R \to {x_l}}},{\gamma _{{D_k} \to {x_l}}}} \right)$, the corresponding CDF $F_X$ is presented in the following lemma.
\begin{lemma}\label{lemma:gamma Bn_CDF}
The CDF $F_X$ for $x_l$ is given by \eqref{the CDF of x1 for ergodic rate} at the top of the next page, where
${f_W}\left( w \right) = \frac{{{{\tilde \lambda }_1}{{\tilde \lambda }_2}}}{{{{\tilde \lambda }_2} - {{\tilde \lambda }_1}}}\left( {{e^{ - {{\tilde \lambda }_1}w}} - {e^{ - {{\tilde \lambda }_2}w}}} \right)$ and ${f_Z}\left( z \right){\rm{ = }}\prod\limits_{i = 1}^3 {{\lambda _i}} \left( {{\Phi _1}{e^{ - {\lambda _1}z}} -
{\Phi _2}{e^{ - {\lambda _2}z}}{\rm{ + }}{\Phi _3}{e^{ - {\lambda _3}z}}} \right)$, ${{\tilde \lambda }_1} = \frac{1}{{\varepsilon \rho }}$, ${{\tilde \lambda }_2} = \frac{1}{{\rho {\varpi _2}}}$. $\varphi {\rm{ = }}\frac{{{a_l}\left( {w + 1} \right){\Omega _l} + {b_l}\left( {z + 1} \right){\Omega _k}}}{{{a_l}\left( {w + 1} \right){\Omega _l}{\Omega _k}}}$ and $\vartheta  = \frac{{{a_l}\left( {w + 1} \right){\Omega _l} + {b_l}\left( {z + 1} \right){\Omega _k}}}{{{b_l}\left( {z + 1} \right){\Omega _k}{\Omega _l}}}$.
\begin{proof}
See Appendix C.
\end{proof}
\end{lemma}
\begin{figure*}[!t]
\normalsize
\begin{align}\label{the CDF of x1 for ergodic rate}
{F_X}\left( x \right) = \int_0^\infty  {\int_0^\infty  {\frac{{{f_W}\left( w \right){f_Z}\left( z \right)}}{{\varphi {\Omega _k}}}} } \left( {1 - {e^{ - \frac{{x\left( {w + 1} \right)\varphi }}{{\rho {b_l}}}}}} \right)dzdw + \int_0^\infty  {\int_0^\infty  {\frac{{{f_W}\left( w \right){f_Z}\left( z \right)}}{{\vartheta {\Omega _l}}}\left( {1 - {e^{ - \frac{{x\left( {z + 1} \right)\vartheta }}{{\rho {a_l}}}}}} \right)} } dzdw.
\end{align}
\hrulefill \vspace*{0pt}
\end{figure*}

Substituting \eqref{the CDF of x1 for ergodic rate}, the corresponding ergodic rate of $x_l$ is given by 
\begin{align}\label{the ergodic rate of x1}
R_{{x_l}}^{erg} = \frac{1}{{2\ln 2}}\int_0^\infty  {\frac{{1 - {F_X}\left( x \right)}}{{1 + x}}} dx,
\end{align}
where $X = \min \left( {{\gamma _{R \to {x_l}}},{\gamma _{{D_k} \to {x_l}}}} \right)$ and $\varepsilon  = 1$. Unfortunately, it is difficult to obtain the closed-form expression from \eqref{the ergodic rate of x1}. However, it can be evaluated by applying numerical approaches. To further obtain analytical results, we consider the special cases of $x_l$ with ipSIC/pSIC for TWR-NOMA where there is no IS between the pair of antennas at the relay in the following part.

Based on the above analysis, for the special case that substituting ${\varpi _1} = {\varpi _2} = 0$ into \eqref{the ergodic rate of x1}, the ergodic rate of $x_l$ with ipSIC can be obtained in the following theorem.
\begin{theorem}\label{the theorem of_ergodic_rate_x1 with ipSIC no IS}
The closed-form expression of ergodic rate for $x_{1}$ with ipSIC for TWR-NOMA is given by
\begin{align}\label{the ergodic rate of x1 with ipSIC no IS}
 R_{{x_l},erg}^{ipSIC} = & \frac{{ - 1}}{{2\ln 2}}\left[ {A{e^\Psi }{{\rm{Ei}}}\left( { - \Psi } \right) + \frac{{B{e^{\frac{\Psi }{{{\Lambda _1}}}}}}}{{{\Lambda _1}}}{{\rm{Ei}} }\left( {\frac{{ - \Psi }}{{{\Lambda _1}}}} \right)} \right. \nonumber \\
 & \left. { + \frac{{C{e^{\frac{\Psi }{{{\Lambda _2}}}}}}}{{{\Lambda _2}}}{{\rm{Ei}}}\left( {\frac{{ - \Psi }}{{{\Lambda _2}}}} \right)} \right] ,
\end{align}
where ${\Lambda _1}{\rm{ = }}\frac{{\varepsilon {\Omega _I}}}{{{b_l}{\Omega _k}}}$, ${\Lambda _2}{\rm{ = }}\frac{{{a_t}{\Omega _t}}}{{{a_l}{\Omega _l}}}$ and $\Psi  = \frac{{{a_l}{\Omega _l} + {b_l}{\Omega _k}}}{{\rho {a_l}{b_l}{\Omega _l}{\Omega _k}}}$;
$A = \frac{1}{{{\Lambda _1}{\Lambda _2} - {\Lambda _2} - {\Lambda _1} + 1}}$,
$B = \frac{{A\left( {{\Lambda _1} - {\Lambda _1}{\Lambda _2}} \right) - {\Lambda _1}}}{{\left( {{\Lambda _2} - {\Lambda _1}} \right)}}$ and $C = 1-A-B$. $\mathrm{Ei\left(\cdot\right)}$ is the exponential integral function~\cite[Eq. (8.211.1)]{gradshteyn}.
\begin{proof}
See Appendix D.
\end{proof}
\end{theorem}

\begin{corollary}
Based on \eqref{the ergodic rate of x1 with ipSIC no IS}, the ergodic rate of $x_{l}$ for pSIC with $\varepsilon = 0$ can be expressed in the closed form as
\begin{align}\label{corollary1 derived for D1 no directlink for HD NOMA }
R_{{x_l},erg}^{pSIC} = \frac{{ - 1}}{{2\ln 2}}\left[ {A{e^\Psi }{{\rm{Ei}}}\left( { - \Psi } \right) + \frac{{C{e^{\frac{\Psi }{{{\Lambda _2}}}}}}}{{{\Lambda _2}}}{{\rm{Ei}} }\left( { - \frac{\Psi }{{{\Lambda _2}}}} \right)} \right].
\end{align}
\end{corollary}
\subsubsection{Ergodic Rate of $x_{t}$}
On the condition that the relay and $D_l$ are capable of detecting $x_t$, $x_t$ can be also detected by $D_t$ successfully. As a consequence, combining \eqref{the expression SINR of R to detect x2 or x4}, \eqref{the SINR expression for D1 or D3 to detect x4 or x2} and \eqref{the SINR expression for D2 or D4}, the achievable rate of $x_t$ is written as ${R_{{x_t}}} = \frac{1}{2}\log \left( {1 + \min \left( {{\gamma _{R \to {x_t}}},{\gamma _{{D_k} \to {x_t}}},{\gamma _{{D_r} \to {x_t}}}} \right)} \right)$. The corresponding ergodic rate of $x_t$ can be expressed as
\begin{align}\label{the ergodic rate of x2}
R_{{x_t}}^{erg} = \frac{1}{{2\ln 2}}\int_0^\infty  {\frac{{1 - {F_Y}\left( y \right)}}{{1 + y}}} dy,
\end{align}
where $Y = \min \left( {{\gamma _{R \to {x_t}}},{\gamma _{{D_k} \to {x_t}}},{\gamma _{{D_r} \to {x_t}}}} \right)$ with ${\varpi _1} = {\varpi _2} = 1$ and $\varepsilon  = 1$. To the best of authors' knowledge, \eqref{the ergodic rate of x2} does not have a closed form solution. We also consider the special cases of $x_t$ by the virtue of ignoring IS between the pair of antennas at the relay.

For the special case that substituting ${\varpi _1} = {\varpi _2} = 0$ into \eqref{the ergodic rate of x2} and after some manipulations, the ergodic rates of $x_t$ with ipSIC/pSIC is given by
\begin{align}\label{the ergodic rate of x2 with ipSIC no IS}
R_{{x_t},erg}^{ipSIC} = \frac{1}{{2\ln 2}}\int_0^{\frac{{{b_t}}}{{{b_l}}}} {\frac{{{e^{ - \frac{x}{{\rho {a_t}{\Omega _t}}} - \frac{x}{{\rho \left( {{b_t} - x{b_l}} \right){\Omega _k}}} - \frac{x}{{\rho \left( {{b_t} - x{b_l}} \right){\Omega _r}}}}}}}{{\left( {1 + x} \right)\left( {1 + x{\Lambda _3}} \right)}}} dx,
\end{align}
and
\begin{align}\label{the ergodic rate of x2 with pSIC no IS}
R_{{x_t},erg}^{pSIC} = \frac{1}{{2\ln 2}}\int_0^{\frac{{{b_t}}}{{{b_l}}}} {\frac{{{e^{ - \frac{x}{{\rho {a_t}{\Omega _t}}} - \frac{x}{{\rho \left( {{b_t} - x{b_l}} \right){\Omega _k}}} - \frac{x}{{\rho \left( {{b_t} - x{b_l}} \right){\Omega _r}}}}}}}{{1 + x}}} dx,
\end{align}
respectively, where ${\Lambda _3}{\rm{ = }}\frac{{\varepsilon {\Omega _I}}}{{{a_t}{\Omega _t}}}$ with $\varepsilon  = 1$.

As can be seen from the above expressions, the exact analysis of ergodic rates require the computation of some complicated integrals. To facilitate these analysis and provide the simpler expression for the ergodic rate of $x_t$ with ipSIC/pSIC, the following theorem and corollary provide the high SNR approximations to evaluate the performance.                                                     \begin{theorem}\label{the theorem of_ergodic_rate_x2 with ipSIC no IS}
The approximation expression for ergodic rate of $x_{t}$ with ipSIC at high SNR is given by
\begin{align}\label{the high SNR approximation for ergodic rate of x2 with ipSIC no IS}
R_{{x_t},\infty }^{ipSIC} = \frac{1}{{2\left( {1 - {\Lambda _3}} \right)\ln 2}}\left[ {\ln \left( {1 + \frac{{{b_t}}}{{{b_l}}}} \right) - \ln \left( {1 + \frac{{{b_t}{\Lambda _3}}}{{{b_l}}}} \right)} \right].
\end{align}
\begin{proof}
See Appendix E.
\end{proof}
\end{theorem}

\begin{corollary}
For the special case with $\varepsilon = 0$, the ergodic rate of $x_{t}$ for pSIC can be approximated at high SNR as
\begin{align}\label{the asymptotic for corollary1 derived for D1 no directlink for HD NOMA }
R_{{x_t},\infty }^{pSIC} = \frac{1}{{2\ln 2}}{e^{\frac{1}{{\rho {a_t}{\Omega _t}}}}}\left[ {{{\rm{Ei}} }\left( {\frac{{ - 1}}{{\rho {a_t}{b_l}{\Omega _t}}}} \right) - {{\rm{Ei}} }\left( {\frac{{ - 1}}{{\rho {a_t}{\Omega _t}}}} \right)} \right].
\end{align}
\end{corollary}
\subsubsection{Slope Analysis}
In this subsection, by the virtue of asymptotic results, we characterize the high SNR slope which is capable of  capturing the influence of channel parameters on the ergodic rate. The high SNR slope is defined as
\begin{align}\label{high SNR slop}
S = \mathop {\lim }\limits_{\rho  \to \infty } \frac{{R_{x_i}^\infty \left( \rho  \right)}}{{\log \left( \rho  \right)}},
\end{align}
where ${R_{x_i}^\infty}$ denotes the asymptotic ergodic rate of $x_{i}$.

\paragraph{$x_{l}$ for ipSIC/pSIC case}
\begin{proposition}\label{The high SNR slop of x1}
Based on the above analytical results in \eqref{the ergodic rate of x1 with ipSIC no IS} and \eqref{corollary1 derived for D1 no directlink for HD NOMA }, when $\rho  \to \infty $, by using ${\mathop{\rm Ei}\nolimits} \left( { - x} \right) \approx \ln \left( x \right) + E_c$~\cite[Eq. (8.212.1)]{gradshteyn} and ${e^{-x}}\approx 1- x$, where $E_c$ is the Euler constant, the asymptotic ergodic rates of $x_l$ with ipSIC/pSIC in the high regime are given by
\begin{align}\label{asymptotic rate of x1 with ipSIC no IS}
&R_{{x_l},\infty }^{ipSIC} = \frac{{ - 1}}{{2\ln 2}}\left[ {A\left( {1 + \Psi } \right)\left( {\ln \left( \Psi  \right) + {E_c}} \right) + \frac{B}{{{\Lambda _1}}}\left( {1 + \frac{\Psi }{{{\Lambda _1}}}} \right)} \right. \nonumber \\
& \left. { \times \left( {\ln \left( {\frac{\Psi }{{{\Lambda _1}}}} \right) + {E_c}} \right) + \frac{{{E_c}}}{{{\Lambda _2}}}\left( {1 + \frac{\Psi }{{{\Lambda _2}}}} \right)\left( {\ln \left( {\frac{\Psi }{{{\Lambda _2}}}} \right) + {E_c}} \right)} \right],
\end{align}
and
\begin{align}\label{asymptotic rate of x1 with pSIC no IS}
R_{{x_l},\infty }^{pSIC} =& \frac{{ - 1}}{{2\ln 2}}\left[ {A\left( {1 + \Psi } \right)\left( {\ln \left( \Psi  \right) + {E_c}} \right)} \right. \nonumber\\
 &\left. { + \frac{{{E_c}}}{{{\Lambda _2}}}\left( {1 + \frac{\Psi }{{{\Lambda _2}}}} \right)\left( {\ln \left( {\frac{\Psi }{{{\Lambda _2}}}} \right) + {E_c}} \right)} \right] ,
\end{align}
respectively.

Substituting \eqref{asymptotic rate of x1 with ipSIC no IS} and \eqref{asymptotic rate of x1 with pSIC no IS} into \eqref{high SNR slop}, we can see that the high SNR slopes of $x_l$ with ipSIC/pSIC are equal to zeros.
\end{proposition}
\paragraph{$x_{t}$ for ipSIC/pSIC case}
Similar to \eqref{asymptotic rate of x1 with ipSIC no IS} and \eqref{asymptotic rate of x1 with pSIC no IS}, substituting \eqref{the high SNR approximation for ergodic rate of x2 with ipSIC no IS} and \eqref{corollary1 derived for D1 no directlink for HD NOMA } into \eqref{high SNR slop}, we observe that the high SNR slopes of $x_t$ with ipSIC/pSIC are also equal to zeros.
\begin{remark}\label{remark:5}
The above analytical results demonstrate that even if there is no IS between both antennas at the relay, $x_l$ and $x_t$  converge to throughput ceilings and obtain zero slopes in the high SNR regime. This is due to the fact that
the first phase is similar to uplink NOMA, it is suffering interference from other users which has seriously impact on the high SNR slope.
\end{remark}
\subsubsection{Throughput Analysis}\label{the delay-tolerant throughput for expression}
In delay-tolerant transmission scenario, the system throughput is determined by evaluating the ergodic rate. Based on the above results derived, the corresponding throughput of TWR-NOMA is given by
\begin{align}\label{delay-tolerant throughput}
R_{dt}^{\psi }  = R_{{x_1},erg}^{\psi}  + R_{{x_2},erg}^{\psi}   + R_{{x_3},erg}^{\psi}   + R_{{x_4},erg}^{\psi},
\end{align}
where $R_{{x_1},erg}^{\psi} $ and $R_{{x_3},erg}^{\psi}$ with ipSIC/PSIC can be obtained from \eqref{the ergodic rate of x1 with ipSIC no IS} and \eqref{corollary1 derived for D1 no directlink for HD NOMA }, respectively, while
$R_{{x_2},erg}^{\psi}$ and $R_{{x_4},erg}^{\psi}$ with ipSIC/pSIC can be obtained from and \eqref{the ergodic rate of x2 with ipSIC no IS}, \eqref{the ergodic rate of x2 with pSIC no IS}, respectively.
\section{Energy Efficiency}\label{Energy Efficiency}
In this section, the performance of TWR-NOMA systems is characterized from the perspective of energy efficiency (EE). In particular, EE has been adopted as a efficient metric to provide quantitative analysis for 5G networks.
The core idea of EE is a rate between the total data rate of all NOMA users and the total energy consumption.
Therefore, the expression of EE can be given by
\begin{align}\label{the defination of EE}
{\eta _{EE}} = \frac{{{\rm{Total~data~rate}}}}{{{\rm{Total~energy~consumption}}}}.
\end{align}

Based on the throughput analysis in \eqref{Seciton-A-3} and \eqref{the delay-tolerant throughput for expression}, the EE of TWR-NOMA systems is given by
\begin{align}\label{EE HD for limited}
\eta _\Upsilon ^{EE} = \frac{{2R_{\Upsilon}^{\psi} }}{{T{P_u} + T{P_r}}},
\end{align}
where $\Upsilon \in \left( {dt,dl} \right)$ and $T$ denotes transmission time of the entire communication process. $\eta _{dl}^{EE}$ and $\eta _{dt}^{EE}$ are the system energy efficiency in delay-limited transmission mode and delay-tolerant transmission mode, respectively.

\begin{table}[!t]
\centering
\caption{Table of Parameters for Numerical Results}
\tabcolsep5pt
\renewcommand\arraystretch{1.1} 
\begin{tabular}{|l|l|}
\hline
Monte Carlo simulations repeated  &  ${10^6}$ iterations \\
\hline
\multirow{2}{*}{Power allocation coefficients of NOMA} &  \multirow{1}{*}{ $b_1=b_3=0.2$}   \\
                                                       &  \multirow{1}{*}{ $b_2=b_4=0.8$}   \\
\hline
\multirow{2}{*}{Targeted data rates}  & \multirow{1}{*}{$R_{{1}}=R_{{3}}=0.1 $ BPCU}  \\
                                      & \multirow{1}{*}{$R_{{2}}=R_{{4}}=0.01$ BPCU}  \\
\cline{1-2}
Pass loss exponent  & $\alpha=2$  \\
\hline
The distance between R and $D_{1}$ or $D_{3}$ &  $d_1=2$ m \\
\hline
The distance between R and $D_{2}$ or $D_{4}$ & $d_2 = 10$ m \\
\hline
\end{tabular}
\label{parameter}
\end{table}
\section{Numerical Results}\label{Numerical Results}
In this section, numerical results are provide to substantiate the system performance and investigate the impact levels of IS on outage probability and ergodic rate for TWR-NOMA. Monte Carlo simulation parameters used are summarized in Table~\ref{parameter}, where BPCU is short for bit per channel use. Due to the reciprocity of channels between user groups (i.e., $G_1$ or $G_2$) and $R$, the outage behaviors and ergodic rates of $x_1$ and $x_2$ in $G_1$ are presented to illustrate availability of TWR-NOMA. Without loss of generality, the power allocation coefficients of $x_1$ and $x_2$ are set as $a_1=0.8$ and $a_2=0.2$, respectively. ${\Omega _1}$ and ${\Omega _2}$ are set to be ${\Omega _1} = d_1^{ - \alpha }$ and ${\Omega _2} = d_2^{ - \alpha }$, respectively. The performance of conventional TWR-OMA is shown as a benchmark for comparison, in which the total communication process can be finished in five slots.
In the first slot, the user nodes  in $G_1$, i,e, $D_1$ and $D_2$ sends signal $x_{1}$ and $x_{2}$ to $R$. Meanwhile, the user nodes in $G_2$, i,e, $D_3$ and $D_4$ sends signal $x_{3}$ and $x_{4}$ to $R$. After completing the exchange of information, $R$ sends signal $x_{3}$ and $x_{4}$ to $D_1$ and $D_2$ in the second and third slots, respectively. Then $R$ sends signal $x_{1}$ and $x_{2}$ to $D_3$ and $D_4$ in the fourth and fifth slots, respectively. Except power allocation coefficients, other simulation parameters of TWR-OMA is similar to that of TWR-NOMA. It is worth pointing out that the signals are transmitted at full power for TWR-OMA.

\subsection{Outage Probability}
\begin{figure}[t!]
    \begin{center}
        \includegraphics[width=3.4in,  height=2.6in]{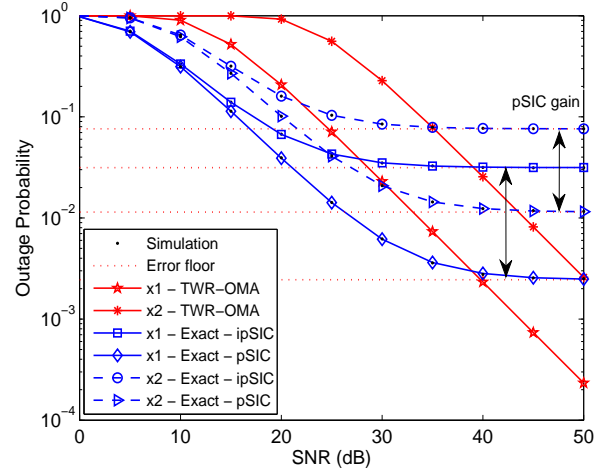}
        \caption{Outage probability versus the transmit SNR, with ${\varpi _1} = {\varpi _2} = 0.01$, ${R_{1}}=0.1$, ${R_{{2}}}=0.01$ BPCU, and ${\Omega _I}=-20$  dB.}
        \label{Pout_x1_x2_ipSIC_pSIC_vs_OMA}
    \end{center}
\end{figure}

Fig. \ref{Pout_x1_x2_ipSIC_pSIC_vs_OMA} plots the outage probabilities of $x_1$ and $x_2$ with both ipSIC and pSIC versus SNR for simulation setting with ${\varpi _1} = {\varpi _2}=0.01$ and ${\Omega _I}=-20$ dB. The solid and dashed curves represent the exact theoretical performance of $x_1$ and $x_2$ for both ipSIC and pSIC, corresponding to the results derived in \eqref{OP derived for x1}, \eqref{corollary1 derived for x1 with perfect SIC} and \eqref{OP derived for x2}, \eqref{corollary2 derived for x2 with perfect SIC}, respectively. Apparently, the outage probability curves match perfectly with Monte Carlo simulation results. As can be observed from the figure, the outage behaviors of $x_1$ and $x_2$ for TWR-NOMA are superior to TWR-OMA in the low SNR regime.
This is due to the fact that the influence of IS is not the dominant factor at low SNR. Hence in this scenario, NOMA systems should work as much as possible at low SNR regime, such as, the wide coverage in rural areas and cell edge scenarios.
Another observation is that the pSIC is capable of enhancing the performance of NOMA compare to the ipSIC.
In addition, the asymptotic curves of $x_1$ and $x_2$ with ipSIC/pSIC are plotted according to \eqref{the asymptotic OP of x1 with ipSIC}, \eqref{the asymptotic OP of x1 with pSIC} and \eqref{the asymptotic OP of x2 with ipSIC}, \eqref{the asymptotic OP of x2 with pSIC}, respectively. It can be seen that the outage behaviors of $x_1$ and $x_2$ converge to the error floors in the high SNR regime. The reason can be explained that due to the impact of residual interference by the use of ipSIC, $x_1$ and $x_2$ result in zero diversity orders. Although the pSIC is carried out in TWR-NOMA system, $x_1$ and $x_2$ also obtain zero diversity orders. This is due to the fact that when the relay first detect the strongest signal in the first slot, it will suffer interference from the weaker signal. This process is similar to the uplink NOMA \cite{TabassumHina2016}. Additionally, this observation verifies the conclusion \textbf{Remark \ref{remark:1}} in Section \ref{Section_III}.
\begin{figure}[t!]
    \begin{center}
        \includegraphics[width=3.4in,  height=2.6in]{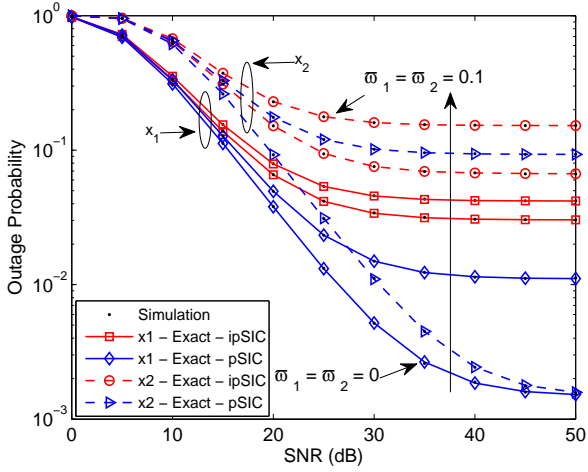}
        \caption{Outage probability versus the transmit SNR, with the different impact levels of IS from ${\varpi _1} = {\varpi _2} = 0$ to ${\varpi _1} = {\varpi _2} = 0.1$, ${R_{1}}=0.1$, ${R_{{2}}}=0.01$ BPCU,
        and ${\Omega _I}=-20$  dB.}
        \label{Pout_x1_x2_ipSIC_pSIC_diff_LI}
    \end{center}
\end{figure}

\begin{figure}[t!]
    \begin{center}
        \includegraphics[width=3.4in,  height=2.6in]{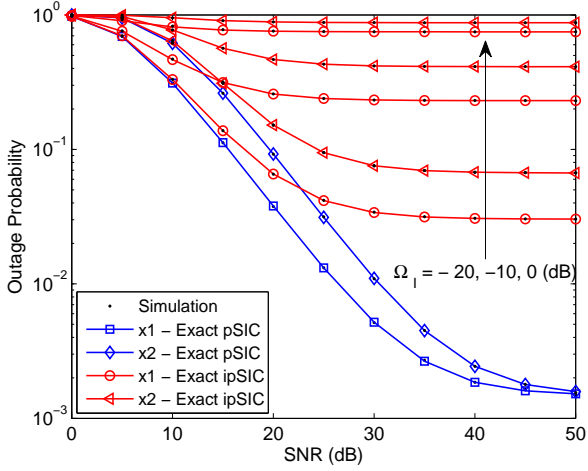}
        \caption{Outage probability versus the transmit SNR, with different values of residual IS from $-20$ dB to $0$ dB,
        ${\varpi _1} = {\varpi _2} = 0$, ${R_{1}}=0.1$, ${R_{{2}}}=0.01$ BPCU.}
        \label{Outage_x1_x2_with_diff_LI_ipSIC}
    \end{center}
\end{figure}

Fig. \ref{Pout_x1_x2_ipSIC_pSIC_diff_LI} plots the outage probabilities of $x_1$ and $x_2$ versus SNR with the different impact levels of IS from ${\varpi _1} = {\varpi _2} = 0$ to ${\varpi _1} = {\varpi _2} = 0.1$.  The solid and dashed curves represent
the outage behaviors of $x_1$ and $x_2$ with ipSIC/pSIC, respectively.
As can be seen that when the impact level of IS is set to be ${\varpi _1} = {\varpi _2} = 0$, there is no IS between $A_1$ and $A_2$ at the relay, which can be viewed as a benchmark. Additionally, one can observed that with the impact levels of IS increasing, the outage performance of TWR-NOMA degrades significantly. As a consequence, it is crucial to hunt for efficient strategies for suppressing the effect of interference between antennas. Fig. \ref{Outage_x1_x2_with_diff_LI_ipSIC} plots the outage probabilities versus SNR with different values of residual IS from $-20$ dB to $0$ dB. It can be seen that the different values of residual IS affects the
performance of ipSIC seriously. Similarly, as the values of residual IS increases, the preponderance of ipSIC is inexistent. When ${\Omega _I}=0$ dB, the outage probabilities of $x_1$ and $x_2$ will be in close proximity to one. Therefore, it is important to design effective SIC schemes for TWR-NOMA.
\begin{figure}[t!]
    \begin{center}
        \includegraphics[width=3.4in,  height=2.6in]{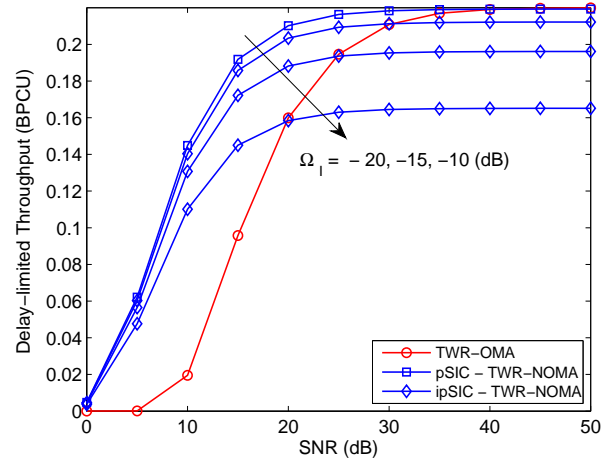}
        \caption{System throughput in delay-limited transmission mode versus SNR with ipSIC/pSIC, ${R_{1}}=0.1$, ${R_{{2}}}=0.01$ BPCU,
        ${\varpi _1} = {\varpi _2} = 0.01$.}
        \label{delay_limited_throughput}
    \end{center}
\end{figure}

Fig. \ref{delay_limited_throughput} plots system throughput versus SNR in delay-limited transmission mode for TWR-NOMA with different values of residual IS from $-20$ dB to $-10$ dB.
The blue solid curves represent throughput for TWR-NOMA with both pSIC and ipSIC, which can be obtained from \eqref{delay-limited throughput}. One can observe that TWR-NOMA is capable of achieving a higher throughput compared to TWR-OMA in the low SNR regime, since it has a lower outage probability. Moreover, the figure confirms that TWR-NOMA converges to the throughput ceiling in the high SNR regime. Additionally, it is worth noting that ipSIC considered for TWR-NOMA will further degrade throughput with the values of residual IS becomes larger in high SNR regimes.
\subsection{Ergodic Rate}
\begin{figure}[t!]
    \begin{center}
        \includegraphics[width=3.4in,  height=2.6in]{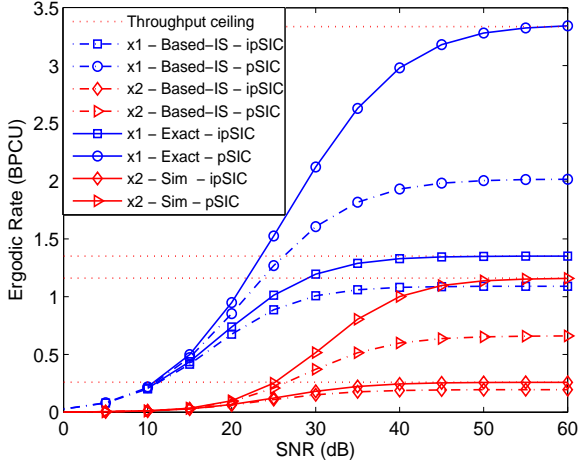}
        \caption{Ergodic rate versus the transmit SNR with ipSIC/pSIC, ${\varpi _1} = {\varpi _2} = 0.01$, and ${\Omega _I}=-20$  dB.}
        \label{erg_achi_rate}
    \end{center}
\end{figure}

Fig. \ref{erg_achi_rate} plots the ergodic rate of $x_1$ and $x_2$ for TWR-NOMA versus SNR and the values of SI are assumed to be ${\varpi _1} = {\varpi _2} = 0.01$, and ${\Omega _I}=-20$  dB.
The blue and red dash-dotted curves represent the achievable rate of $x_1$ and $x_2$ with ipSIC/pSIC for TWR-NOMA, respectively, which considers IS between $A_1$ and $A_2$ at the relay. The blue and red solid curves represent ergodic rates of $x_1$ and $x_2$ with ipSIC/pSIC according to \eqref{the ergodic rate of x1 with ipSIC no IS}, \eqref{corollary1 derived for D1 no directlink for HD NOMA } and \eqref{the ergodic rate of x2 with ipSIC no IS}, \eqref{the ergodic rate of x2 with pSIC no IS}, respectively.
We can observe that the ergodic rates of $x_1$ and $x_2$ with pSIC are larger than that of $x_1$ and $x_2$ with ipSIC.
This is due to the fact that pSIC can provide more performance gain than ipSIC.
In addition, due to the influence of interference, $x_1$ and $x_2$ converge to the throughput ceilings in high SNR regimes, which verifies the conclusion \textbf{Remark \ref{remark:5}} in Section \ref{Ergodic rate}.

\begin{figure}[t!]
    \begin{center}
        \includegraphics[width=3.4in,  height=2.6in]{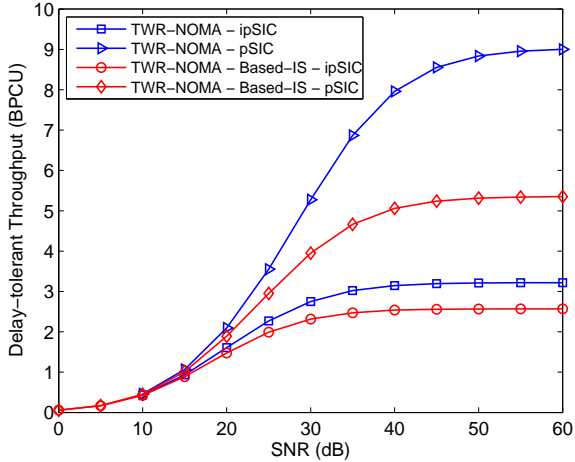}
        \caption{System throughput in delay-tolerant transmission mode versus SNR with ipSIC/pSIC,
        ${\varpi _1} = {\varpi _2} = 0.01$, and ${\Omega _I}=-20$  dB.}
         \label{delay_toletant_throughput}
    \end{center}
\end{figure}

Fig. \ref{delay_toletant_throughput} plots the system throughput versus SNR in delay-tolerant transmission mode for TWR-NOMA. The blue solid curves represent system throughput for TWR-NOMA with ipSIC/pSIC, which can be obtained from \eqref{delay-limited throughput}. The system throughput of IS-based is selected to be the benchmark denoted by the red dash-dotted curves. It is observed that TWR-NOMA can achieve a higher throughput in the absence of IS at the relay. Hence, we need to find an effective way to restrain IS for both antennas at the relay.
\begin{figure}[t!]
    \begin{center}
        \includegraphics[width=3.4in,  height=2.6in]{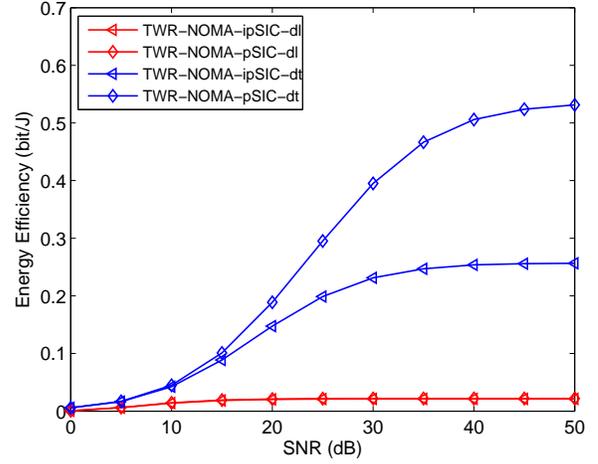}
        \caption{System throughput in delay-limited/tolerant transmission mode versus SNR with ipSIC/pSIC, where $P_{u}=P_{r}=$10 W, and $T=1$.
        }
         \label{Energy Efficiency for dl and dt}
    \end{center}
\end{figure}
\subsection{Energy Efficiency}
Fig. \ref{Energy Efficiency for dl and dt} plots energy efficiency of TWR-NOMA systems versus SNR with delay-limited/tolerant transmission modes.
The red solid curves represent system energy efficiency for the delay-limited transmission mode with ipSIC/pSIC, respectively, which can be obtained from \eqref{delay-limited throughput} and \eqref{EE HD for limited}.
The blue curves represent system energy efficiency for the delay-tolerant transmission mode with ipSIC/pSIC, respectively, which can be obtained from \eqref{delay-tolerant throughput} and \eqref{EE HD for limited}. It is can be seen that TWR-NOMA with ipSIC/pSIC in delay-limited transmission mode have almost the same energy efficiency. Additionally, we can observed that the energy efficiency of TWR-NOMA with pSIC is superior to ipSIC in high SNR regimes.
\section{Conclusion}\label{Conclusion}
This paper has investigated the application of TWR to NOMA systems, in which two pairs of users can exchange their information between each other by the virtue of a relay node. The performance of TWR-NOMA systems has been characterized in terms of outage probability and ergodic rate for both ipSIC and pSIC. The closed-form expressions of outage probability for the NOMA users' signals have been derived. Owing to the impact of IS at relay, there were the error floors for TWR-NOMA with ipSIC/pSIC in high SNR regimes and zero diversity orders were obtained. Based on the analytical results, it was shown that the performance of TWR-NOMA with ipSIC/pSIC outperforms TWR-OMA in the low SNR regime. Furthermore, the ergodic rates of TWR-NOMA have been discussed in detail. The results have shown that TWR-NOMA with pSIC is capable of achieving a larger rate in the absence of IS at the relay. More particularlly, the users' signals for TWR-NOMA converge to the throughput ceiling and gain zero high slopes in high SNR regimes.
Finally, the system energy efficiencies with ipSIC/pSIC were discussed in a pair of transmission
modes.
\appendices
\section*{Appendix~A: Proof of Theorem \ref{theorem:1 the outage of x1}} \label{Appendix:A}
\renewcommand{\theequation}{A.\arabic{equation}}
\setcounter{equation}{0}

Substituting \eqref{the expression SINR of R to detect x1 or x3}, \eqref{the SINR expression for D1 or D3 to detect x4 or x2} and
\eqref{the SINR expression for D1 or D3 to detect its own information} into \eqref{OP expression for x1}, the outage probability of $x_l$
can be further given by
\begin{align*}\label{the proof of outage probability for x1 with imperfect SIC}
 &P_{{x_l}}^{ipSIC} = 1 \nonumber \\
  \end{align*}
 \begin{align}
  &- \underbrace {\Pr \left( {\frac{{\rho {{\left| {{h_l}} \right|}^2}{a_l}}}{{\rho {{\left| {{h_t}} \right|}^2}{a_t} + \rho {\varpi _1}({{\left| {{h_k}} \right|}^2}{a_k} + {{\left| {{h_r}} \right|}^2}{a_r}) + 1}} > {\gamma _{t{h_l}}}} \right)}_{{J_1}}\nonumber \\
 & \times \Pr \left( {\frac{{\rho {{\left| {{h_k}} \right|}^2}{b_t}}}{{\rho {{\left| {{h_k}} \right|}^2}{b_l} + \rho {\varpi _2}{{\left| {{h_k}} \right|}^2} + 1}} > {\gamma _{t{h_t}}},} \right. \nonumber\\
&\underbrace {\begin{array}{*{20}{c}}
   {} & {} & {} & {} & {} & {}  \\
\end{array}\left. {\frac{{\rho {{\left| {{h_k}} \right|}^2}{b_l}}}{{\varepsilon \rho {{\left| g \right|}^2} + \rho {\varpi _2}{{\left| {{h_k}} \right|}^2} + 1}} > {\gamma _{t{h_l}}}} \right)}_{{J_2}}  ,
\end{align}
where $\varepsilon =1$.

To calculate the probability $J_1$ in \eqref{the proof of outage probability for x1 with imperfect SIC},
let $Z = \rho {a_t}{\left| {{h_t}} \right|^2} + \rho {\varpi _1}{a_k}{\left| {{h_k}} \right|^2} + \rho {\varpi _1}{a_r}{\left| {{h_r}} \right|^2}$. We first calculate the PDF of $Z$ and then give the process derived of $J_{1}$. As is known, ${\left| {{h_i}}
\right|^2}$ follows the exponential distribution with the parameters ${\Omega _i}$, $i \in \left( {1,2,3,4} \right)$. Furthermore, we denote that ${Z_1} = \rho {a_t}{\left| {{h_t}} \right|^2}$, ${Z_2} = \rho {\varpi _1}{a_k}{\left| {{h_k}} \right|^2}$ and ${Z_3} = \rho {\varpi _1}{a_r}{\left| {{h_r}} \right|^2}$ are also independent exponentially distributed random variables (RVs) with parameters ${\lambda _1}{\rm{ = }}\frac{1}{{\rho {a_t}{\Omega _t}}}$, ${\lambda _2}{\rm{ = }}\frac{1}{{\rho {\varpi _1}{a_k}{\Omega _k}}}$ and ${\lambda _3}{\rm{ = }}\frac{1}{{\rho {\varpi _1}{a_r}{\Omega _r}}}$, respectively. Based on \cite{Nadarajah6831670}, for the independent non-identical distributed (i.n.d) fading
scenario, the PDF of $Z$ can be given by
\begin{align}\label{PDF of Z}
{f_Z}\left( z \right){\rm{ = }}\prod\limits_{i = 1}^3 {{\lambda _i}} \left( {{\Phi _1}{e^{ - {\lambda _1}z}} -
{\Phi _2}{e^{ - {\lambda _2}z}}{\rm{ + }}{\Phi _3}{e^{ - {\lambda _3}z}}} \right),
\end{align}
where ${\Phi _1}{\rm{ = }}\frac{1}{{\left( {{\lambda _2} - {\lambda _1}} \right)\left( {{\lambda _3} - {\lambda _1}}
\right)}}$, ${\Phi _2}{\rm{ = }}\frac{1}{{\left( {{\lambda _3} - {\lambda _2}} \right)\left( {{\lambda _2} - {\lambda _1}} \right)}}$ and ${\Phi _3}{\rm{ = }}\frac{1}{{\left( {{\lambda _3} - {\lambda _1}} \right)\left( {{\lambda _3} - {\lambda _2}} \right)}}$.

According to the above explanations, $J_1$ is calculated as follows:
\begin{align}\label{derived process for J1}
{J_1} = \Pr \left( {{{\left| {{h_l}} \right|}^2} > \left( {Z + 1} \right)\beta_l } \right) = \int_0^\infty  {{f_Z}\left( z \right)} {e^{ - \frac{{\left( {z + 1} \right)\beta_l }}{{{\Omega _l}}}}}dz,
\end{align}
where $\beta_l {\rm{ = }}\frac{{{\gamma _{t{h_l}}}}}{{\rho {a_l}}}$.
Substituting \eqref{PDF of Z} into \eqref{derived process for J1} and after some algebraic manipulations, $J_1$ is given by
\begin{align}\label{the derived expression of J1 finally}
{J_1} = {e^{ - \frac{\beta_l }{{{\Omega _l}}}}}\prod\limits_{i = 1}^3 {{\lambda _i}} \left( {\frac{{{\Phi _1}{\Omega _l}}}{{{\Omega _l}{\lambda _1}{\rm{ + }}\beta_l }} - \frac{{{\Phi _2}{\Omega _l}}}{{{\Omega _l}{\lambda _2}{\rm{ + }}\beta_l }} + \frac{{{\Phi _3}{\Omega _l}}}{{{\Omega _l}{\lambda _3}{\rm{ + }}\beta_l }}} \right),
\end{align}

$J_2$ can be further calculated as follows:
\begin{align}\label{the derived process of J2}
 {J_2} =& \Pr \left( {{{\left| {{h_k}} \right|}^2} > \xi_t ,{{\left| g \right|}^2} < \frac{{{{\left| {{h_k}} \right|}^2} - \tau_l }}{{\varepsilon \rho \tau_l }},{{\left| {{h_k}} \right|}^2} > \tau_l } \right) \nonumber  \\
  =& \Pr \left( {{{\left| {{h_k}} \right|}^2} > \max \left( {\tau_l ,\xi_t } \right) \buildrel \Delta \over = \theta_l ,{{\left| g \right|}^2} < \frac{{{{\left| {{h_k}} \right|}^2} - \tau_l }}{{\varepsilon \rho \tau_l }}} \right) \nonumber\\
  =& \int_\theta ^\infty  {\frac{1}{{{\Omega _k}}}} \left( {{e^{ - \frac{y}{{{\Omega _k}}}}} - {e^{ - \frac{{y - \tau_l }}{{\varepsilon \tau_l \rho {\Omega _I}}} - \frac{y}{{{\Omega _k}}}}}} \right)dy \nonumber\\
  =& {e^{ - \frac{\theta_l }{{{\Omega _k}}}}} - \frac{{\tau_l \varepsilon \rho {\Omega _I}}}{{{\Omega _k} + \varepsilon \rho \tau_l {\Omega _I}}}{e^{ - \frac{{\theta_l \left( {{\Omega _k} + \rho \tau_l \varepsilon {\Omega _I}} \right)}}{{\tau_l \varepsilon \rho {\Omega _I}{\Omega _k}}}{\rm{ + }}\frac{1}{{\varepsilon \rho {\Omega _I}}}}} ,
\end{align}
where $\xi_t {\rm{ = }}\frac{{{\gamma _{t{h_t}}}}}{{\rho \left( {{b_t} - {b_l}{\gamma _{t{h_t}}} - {\varpi _2}{\gamma _{t{h_t}}}} \right)}}$ with ${b_t} > \left( {{b_l} + {\varpi _2}} \right){\gamma _{t{h_t}}}$, $\tau_l {\rm{ = }}\frac{{{\gamma _{t{h_l}}}}}{{\rho \left( {{b_l} - {\varpi _2}{\gamma _{t{h_l}}}} \right)}}$ with ${b_l} > {\varpi _2}{\gamma _{t{h_l}}}$.

Combining \eqref{the derived expression of J1 finally} and \eqref{the derived process of J2}, we can obtain \eqref{OP derived for x1}.

The proof is complete.

\appendices
\section*{Appendix~B: Proof of Theorem \ref{theorem:2 the outage of x2}} \label{Appendix:B}
\renewcommand{\theequation}{B.\arabic{equation}}
\setcounter{equation}{0}

Substituting \eqref{the expression SINR of R to detect x1 or x3}, \eqref{the expression SINR of R to detect x2 or x4}, \eqref{the SINR expression for D1 or D3 to detect its own information} and \eqref{the SINR expression for D2 or D4} into \eqref{OP expression for x2}, the outage probability of $x_t$ is rewritten as
\begin{align}\label{the proof of outage probability for x2 with perfect SIC}
&P_{{x_t}}^{ipSIC} = 1 \nonumber \\
&- \Pr \left( {\frac{{\rho {{\left| {{h_t}} \right|}^2}{a_t}}}{{\varepsilon \rho {{\left| g \right|}^2} + \rho {\varpi _1}({{\left| {{h_k}} \right|}^2}{a_k} + {{\left| {{h_r}} \right|}^2}{a_r}) + 1}} > {\gamma _{t{h_t}}},} \right.\nonumber \\
& \begin{array}{*{20}{c}}
   {}  \\
\end{array}\underbrace {\left. {\frac{{\rho {{\left| {{h_l}} \right|}^2}{a_l}}}{{\rho {{\left| {{h_t}} \right|}^2}{a_t} + \rho {\varpi _1}({{\left| {{h_k}} \right|}^2}{a_k} + {{\left| {{h_r}} \right|}^2}{a_r}) + 1}} > {\gamma _{t{h_l}}}} \right)}_{{\Theta _1}} \nonumber\\
 & \times \underbrace {\Pr \left( {\frac{{\rho {{\left| {{h_k}} \right|}^2}{b_t}}}{{\rho {{\left| {{h_k}} \right|}^2}{b_l} + \rho {\varpi _2}{{\left| {{h_k}} \right|}^2} + 1}} > {\gamma _{t{h_t}}}} \right)}_{{\Theta _2}} \nonumber\\
 & \times \underbrace {\Pr \left( {\frac{{\rho {{\left| {{h_r}} \right|}^2}{b_t}}}{{\rho {{\left| {{h_r}} \right|}^2}{b_l} + \rho {\varpi _2}{{\left| {{h_r}} \right|}^2} + 1}} > {\gamma _{t{h_t}}}} \right)}_{{\Theta _3}} ,
\end{align}
where ${\varpi _1} = {\varpi _2} \in \left[ {0,1} \right]$  and $\varepsilon  = 1$.

Similar to \eqref{PDF of Z}, let ${Z^{'}}{\rm{ = }}\rho {\varpi _1}{a_k}{\left| {{h_k}} \right|^2} + \rho {\varpi _1}{a_r}{\left| {{h_r}} \right|^2}$,  the PDF of ${{Z^{'}}}$ is given by
\begin{align}\label{PDF of Z with two varibles}
{f_{{Z^{'}}}}\left( {{z}}^{'} \right) = \prod\limits_{i = 1}^2 {\lambda _i^{'}} \left( {\frac{{{e^{ - \lambda _1^{'}{{z^{'}}}}}}}{{\left( {\lambda _2^{'} - \lambda _1^{'}}
 \right)}} - \frac{{{e^{ - \lambda _2^{'}{{z^{'}}}}}}}{{\left( {\lambda _2^{'} - \lambda _1^{'}} \right)}}} \right),
\end{align}
where $\lambda _1^{'}{\rm{ = }}\frac{1}{{\rho {\varpi _1}{a_k}{\Omega _k}}}$ and $\lambda _2^{'}{\rm{ = }}\frac{1}
{{\rho {\varpi _1}{a_r}{\Omega _r}}}$.

After some variable substitutions and manipulations,
\begin{align}\label{the derived process of Theta1}
 {\Theta _1} =& \Pr \left( {{{\left| {{h_t}} \right|}^2} > {\beta_t}\left( {\varepsilon \rho {{\left| g \right|}^2} + {Z^{'}} + 1} \right),} \right. \nonumber \\
 &\begin{array}{*{20}{c}}
   {} & {} & {} & {} & {\left. {{{\left| {{h_l}} \right|}^2} > \beta_l \left( {\rho {{\left| {{h_t}} \right|}^2}{a_t} + {Z^{'}} + 1} \right)} \right)}  \nonumber \\
\end{array} \nonumber \\
  = & \int_0^\infty  {{f_{Z^{'}}}\left( z^{'} \right){e^{ - \frac{{\beta_l \left( {z^{'} + 1} \right)}}{{{\Omega _l}}}}}} \nonumber \\
 & \times \int_0^\infty  {{f_{{{\left| g \right|}^2}}}\left( y \right)} \frac{1}{{\varphi_t {\Omega _t}}}{e^{ - {\beta_t}\left( {\varepsilon \rho y  + z^{'} + 1} \right)\varphi_t }}dydz^{'}   \nonumber \\
  =& \frac{1}{{\varphi_t {\Omega _t}\left( {1 + \varepsilon \rho{\beta_t}    \varphi_t {\Omega _I}} \right)}}{e^{ - \frac{\beta_l }{{{\Omega _1}}} - {\beta_t}\varphi_t }} \nonumber\\
  &\times \int_0^\infty  {{f_{Z^{'}}}\left( z^{'} \right){e^{ - \frac{{\left( {\beta_l  + {\beta_t}{\Omega _l}\varphi_t } \right)z^{'}}}{{{\Omega _l}}}}}} dz^{'} ,
\end{align}
where ${\beta_t} = \frac{{{\gamma _{t{h_t}}}}}{{\rho {a_t}}}$ and $\varphi_t  = \frac{{{\Omega _l} + \rho \beta_l {a_t}{\Omega _t}}}
{{{\Omega _l}{\Omega _t}}}$.

Substituting \eqref{PDF of Z with two varibles} into \eqref{the derived process of Theta1}, ${\Theta _1}$ can be given by
\begin{align}\label{the expression of Theta1}
 &{\Theta _1} = \frac{{{e^{ - \frac{\beta_l }{{{\Omega _l}}} - {\beta_t}\varphi_t }}}}{{\varphi_t {\Omega _t}\left( {1 + {\beta_t}\varepsilon \rho \varphi_t {\Omega _I}} \right)\left( {\lambda _2^{'} - \lambda _1^{'}} \right)}} \nonumber \\
  &\times \prod\limits_{i = 1}^2 {\lambda _i^{'}} \left( {\frac{{{\Omega _l}}}{{\beta_l  + {\beta_t}{\Omega _l}\varphi_t  + {\Omega _l}\lambda _1^{'}}} - \frac{{{\Omega _l}}}{{\beta_l  + {\beta_t}{\Omega _l}\varphi_t  + {\Omega _l}\lambda _2^{'}}}} \right) .
\end{align}

${\Theta _2}$ and ${\Theta _3}$ can be easily calculated
\begin{align}\label{the expression of Theta2}
 {\Theta _2} = \Pr \left( {{{\left| {{h_k}} \right|}^2} > \xi_t } \right){\rm{ = }}{e^{ - \frac{\xi_t }{{{\Omega _k}}}}},
\end{align}
and
\begin{align}\label{the expression of Theta3}
{\Theta _3} = \Pr \left( {{{\left| {{h_r}} \right|}^2} > \xi_t } \right) = {e^{ - \frac{\xi_t }{{{\Omega _r}}}}},
\end{align}
respectively, where $\xi_t {\rm{ = }}\frac{{{\gamma _{t{h_t}}}}}{{\rho \left( {{b_t} - {b_l}{\gamma _{t{h_t}}} - {\varpi _2}{\gamma _{t{h_t}}}} \right)}}$ with ${b_t} > \left( {{b_l} + {\varpi _2}} \right){\gamma _{t{h_t}}}$.

Finally, combining \eqref{the expression of Theta1}, \eqref{the expression of Theta2} and \eqref{the expression of Theta3}, we can obtain
\eqref{OP derived for x2} and the proof is completed.

\numberwithin{equation}{section}
\section*{Appendix~C: Proof of Lemma~\ref{lemma:gamma Bn_CDF}} \label{Appendix:C}
\renewcommand{\theequation}{C.\arabic{equation}}
\setcounter{equation}{0}
To derive the CDF $F_X$, based on \eqref{the expression SINR of R to detect x1 or x3} and \eqref{the SINR expression for D1 or D3 to detect its own information}, we can formulate
\begin{align}\label{the first derived process of X}
 {F_X}\left( x \right) =& \Pr \left( {\min \left( {\frac{{\rho {{\left| {{h_l}} \right|}^2}{a_l}}}{{Z + 1}},\frac{{\rho {{\left| {{h_k}} \right|}^2}{b_l}}}{{W + 1}}} \right) < x} \right),\nonumber \\
  =& \underbrace {{\Pr }\left( {\frac{{\rho {{\left| {{h_k}} \right|}^2}{b_l}}}{{W + 1}} < \frac{{\rho {{\left| {{h_l}} \right|}^2}{a_l}}}{{Z + 1}},\frac{{\rho {{\left| {{h_k}} \right|}^2}{b_l}}}{{W + 1}} < x} \right)}_{{Q_1}}\nonumber \\
  &+ \underbrace {{\Pr }\left( {\frac{{\rho {{\left| {{h_l}} \right|}^2}{a_l}}}{{Z + 1}} < \frac{{\rho {{\left| {{h_k}} \right|}^2}{b_l}}}{{W + 1}},\frac{{\rho {{\left| {{h_l}} \right|}^2}{a_l}}}{{Z + 1}} < x} \right)}_{{Q_2}} ,
\end{align}
where $Z = \rho {a_t}{\left| {{h_w}} \right|^2} + \rho {a_k}{\varpi _1}{\left| {{h_k}} \right|^2} + \rho {a_r}{\varpi _1}{\left| {{h_r}} \right|^2}$ and $W=\varepsilon \rho {\left| g \right|^2} + \rho {\varpi _2}{\left| {{h_k}} \right|^2}$. For the i.n.d variable, based on \eqref{PDF of Z} and \eqref{PDF of Z with two varibles}, the PDF $f_Z$ and $f_W$ can be written as
${f_Z}\left( z \right){\rm{ = }}\prod\limits_{i = 1}^3 {{\lambda _i}} \left( {{\Phi _1}{e^{ - {\lambda _1}z}} -
{\Phi _2}{e^{ - {\lambda _2}z}}{\rm{ + }}{\Phi _3}{e^{ - {\lambda _3}z}}} \right)$ and ${f_W}\left( w \right) = \frac{{{{\tilde \lambda }_1}{{\tilde \lambda }_2}}}{{{{\tilde \lambda }_2} - {{\tilde \lambda }_1}}}\left( {{e^{ - {{\tilde \lambda }_1}w}} - {e^{ - {{\tilde \lambda }_2}w}}} \right)$, respectively, where ${{\tilde \lambda }_1} = \frac{1}{{\varepsilon \rho }}$ and ${{\tilde \lambda }_2} = \frac{1}{{\rho {\varpi _2}}}$.

$Q_1$ can be calculated as follows:
\begin{align}\label{the derived process of Q1}
{Q_1} =&\Pr \left( {{{\left| {{h_l}} \right|}^2} > \frac{{{{\left| {{h_k}} \right|}^2}{b_l}\left( {Z + 1} \right)}}{{{a_l}\left( {W + 1} \right)}},{{\left| {{h_k}} \right|}^2} < \frac{{x\left( {W + 1} \right)}}{{\rho {b_l}}}} \right) \nonumber \\
   = & \int_0^\infty  {\int_0^\infty  {{f_W}\left( w \right){f_Z}\left( z \right)} } \int_0^{\frac{{x\left( {w + 1} \right)}}{{\rho {b_l}}}} {\frac{{  {e^{ - u\varphi }}}}{{{\Omega _k}}}} dudzdt \nonumber \\
  = & \int_0^\infty  {\int_0^\infty  {\frac{{{f_W}\left( w \right){f_Z}\left( z \right)}}{{\varphi {\Omega _k}}}} } \left( {1 - {e^{ - \frac{{x\left( {w + 1} \right)\varphi }}{{\rho {b_l}}}}}} \right)dzdw,
\end{align}
where $\varphi {\rm{ = }}\frac{{{a_l}\left( {w + 1} \right){\Omega _l} + {b_l}\left( {z + 1} \right){\Omega _k}}}{{{a_l}\left( {w + 1} \right){\Omega _l}{\Omega _k}}}$.

Similar to \eqref{the derived process of Q1}, after some algebraic manipulations, $Q_2$ is given by
\begin{align}\label{the derived process of Q2}
{Q_2}  =\int_0^\infty  {\int_0^\infty  {\frac{{{f_W}\left( w \right){f_Z}\left( z \right)}}{{\vartheta {\Omega _l}}}\left( {1 - {e^{ - \frac{{x\left( {z + 1} \right)\vartheta }}{{\rho {a_l}}}}}} \right)} } dzdw,
\end{align}
where $\vartheta  = \frac{{{a_l}\left( {w + 1} \right){\Omega _l} + {b_l}\left( {z + 1} \right){\Omega _k}}}{{{b_l}\left( {z + 1} \right){\Omega _k}{\Omega _l}}}$.

Combine \eqref{the derived process of Q1} and \eqref{the derived process of Q2}, we can obtain \eqref{the CDF of x1 for ergodic rate}.

 The proof is completed.

\appendices
\section*{Appendix~D: Proof of Theorem \ref{the theorem of_ergodic_rate_x1 with ipSIC no IS}} \label{Appendix:D}
\renewcommand{\theequation}{D.\arabic{equation}}
\setcounter{equation}{0}
The proof starts by substituting ${\varpi _1} = {\varpi _2} = 0$ into \eqref{the ergodic rate of x1}, the ergodic rate of $x_l$ with ipSIC is given by
\begin{align}\label{the proof of x1 with ipSIC no IS}
R_{{x_l},erg}^{ipSIC} =& {\frac{1}{2}}\mathbb{E} \left[ {  \log \left( {1 + \underbrace {\min \left( {\frac{{\rho {{\left| {{h_k}} \right|}^2}{b_l}}}{{\varepsilon \rho {{\left| g \right|}^2} + 1}},\frac{{\rho {{\left| {{h_l}} \right|}^2}{a_l}}}{{\rho {{\left| {{h_t}} \right|}^2}{a_t} + 1}}} \right)}_U} \right)} \right] \nonumber \\
 =& \frac{1}{{2\ln 2}}\int_0^\infty  {\frac{{1 - {F_U}\left( u \right)}}{{1 + u}}} du,
\end{align}
where $\varepsilon  = 1$.

Applying some algebraic manipulations, the CDF of $U$ can be given by
\begin{align}\label{the CDF of x1 with ipSIC no IS}
{F_U}\left( u \right) = 1 - \frac{{{e^{ - u\Psi }}}}{{\left( {1 + u{\Lambda _1}} \right)\left( {1 + u{\Lambda _2}} \right)}},
\end{align}
where ${\Lambda _1}{\rm{ = }}\frac{{\varepsilon {\Omega _I}}}{{{b_l}{\Omega _k}}}$, ${\Lambda _2}{\rm{ = }}\frac{{{a_t}{\Omega _t}}}{{{a_l}{\Omega _l}}}$ and $\Psi  = \frac{{{a_l}{\Omega _l} + {b_l}{\Omega _l}}}{{\rho {a_l}{b_l}{\Omega _l}{\Omega _k}}}$.

Substituting \eqref{the CDF of x1 with ipSIC no IS} into \eqref{the proof of x1 with ipSIC no IS}, the ergodic rate of $x_l$ with ipSIC can be further expressed as follows:
\begin{align}\label{the final expression for ergodic rate of x1 with ipSIC no IS}
 R_{{x_l},erg}^{ipSIC} =& \frac{1}{{2\ln 2}}\int_0^\infty  {\frac{{{e^{ - u\Psi }}}}{{\left( {1 + u} \right)\left( {1 + u{\Lambda _1}} \right)\left( {1 + u{\Lambda _2}} \right)}}} du \nonumber \\
= & \frac{1}{{2\ln 2}}\int_0^\infty \left( {  {\frac{{A{e^{ - u\Psi }}}}{{1 + u}} + \frac{{B{e^{ - u\Psi }}}}{{1 + u{\Lambda _1}}} + \frac{{C{e^{ - u\Psi }}}}{{1 + u{\Lambda _2}}}}  } \right) du \nonumber\\
  =& \frac{{ - 1}}{{2\ln 2}}\left[ {A{e^\Psi }{{\rm{Ei}} }\left( { - \Psi } \right) + \frac{{B{e^{\frac{\Psi }{{{\Lambda _1}}}}}}}{{{\Lambda _1}}}{{\rm{Ei}} }\left( {\frac{{ - \Psi }}{{{\Lambda _1}}}} \right)} \right.\nonumber \\
 &\left. { + \frac{{C{e^{\frac{\Psi }{{{\Lambda _2}}}}}}}{{{\Lambda _2}}}{{\rm{Ei}} }\left( {\frac{{ - \Psi }}{{{\Lambda _2}}}} \right)} \right],
\end{align}
where $A = \frac{1}{{{\Lambda _1}{\Lambda _2} - {\Lambda _2} - {\Lambda _1} + 1}}$, $B = \frac{{A\left( {{\Lambda _1} - {\Lambda _1}{\Lambda _2}} \right) - {\Lambda _1}}}{{\left( {{\Lambda _2} - {\Lambda _1}} \right)}}$ and $C = 1 - A - B$; \eqref{the final expression for ergodic rate of x1 with ipSIC no IS} can be obtained by using \cite[Eq. (3.352.4)]{gradshteyn}.

The proof is completed.

\section*{Appendix~E: Proof of Theorem \ref{the theorem of_ergodic_rate_x2 with ipSIC no IS}} \label{Appendix:D}
\renewcommand{\theequation}{E.\arabic{equation}}
\setcounter{equation}{0}

We can rewrite \eqref{the ergodic rate of x2 with ipSIC no IS} as follows:
\begin{align}\label{the high SNR approximation of x2 with ipSIC no IS}
R_{{x_t},erg}^{ipSIC} = \frac{1}{2}\mathbb{E}\left[ \log \left( {1 + \underbrace {\min \left( \begin{array}{l}
 \frac{{\rho {{\left| {{h_t}} \right|}^2}{a_t}}}{{\varepsilon \rho {{\left| g \right|}^2} + 1}},\frac{{\rho {{\left| {{h_k}} \right|}^2}{b_t}}}{{\rho {{\left| {{h_k}} \right|}^2}{b_l} + 1}}, \\
 \frac{{\rho {{\left| {{h_r}} \right|}^2}{b_t}}}{{\rho {{\left| {{h_r}} \right|}^2}{b_l} + 1}}
 \end{array} \right)}_{{{Q_1}}}} \right) \right],
\end{align}
where $\varepsilon = 1$.

At high SNR regime, ${Q_1}$ can be approximated as
\begin{align}\label{Q1}
{Q_1} \approx \underbrace {\min \left( {\frac{{{{\left| {{h_t}} \right|}^2}{a_t}}}{{\varepsilon {{\left| g \right|}^2}}},\frac{{{b_t}}}{{{b_l}}}} \right)}_X.
\end{align}
As such, the CDF of X in \eqref{Q1} can be given by
\begin{align}\label{the CDF of X for Q1}
{F_X}\left( x \right){\rm{ = }}1 - \frac{1}{{1 + x{\Lambda _3}}},0 < x < \frac{{{b_t}}}{{{b_l}}},
\end{align}
where ${\Lambda _3}{\rm{ = }}\frac{{\varepsilon {\Omega _I}}}{{{a_t}{\Omega _t}}}$. Substituting \eqref{the CDF of X for Q1} into \eqref{the high SNR approximation of x2 with ipSIC no IS} and  through some manipulations, the approximation solution for ergodic rate of $x_t$ with ipSIC at the high SNR regime can be obtained in \eqref{the high SNR approximation for ergodic rate of x2 with ipSIC no IS}.

The proof is completed.

\bibliographystyle{IEEEtran}
\bibliography{mybib}

\end{document}